\documentclass[manuscript,screen]{acmart}

\AtBeginDocument{%
  \providecommand\BibTeX{{%
    \normalfont B\kern-0.5em{\scshape i\kern-0.25em b}\kern-0.8em\TeX}}}

\usepackage{amsmath,amsfonts}
\usepackage{graphicx}
\usepackage{textcomp}
\usepackage{color}
\usepackage{colortbl}
\usepackage{mathtools}
\usepackage{algorithm}
\usepackage{algpseudocode}
\usepackage{multirow}
\usepackage{soul}
\usepackage{listings}
\usepackage{tcolorbox}
\usepackage{tabularx}
\usepackage{adjustbox}
\usepackage{listings}

\usepackage{xcolor}
\definecolor{dkgreen}{rgb}{0,0.6,0}
\definecolor{gray}{rgb}{0.5,0.5,0.5}
\definecolor{mauve}{rgb}{0.58,0,0.82}
\usepackage{breakcites}

\usepackage{enumitem}
\setlist[itemize]{noitemsep, topsep=0pt}
\usepackage[draft,inline,nomargin,index]{fixme}
\fxsetup{theme=color,mode=multiuser}
\FXRegisterAuthor{Ricky}{asv}{\color{blue}}

\newcommand{\chen}[1]{\textcolor{red}{#1}}

\newcommand{\specialcell}[2][c]{%
  \begin{tabular}[#1]{@{}c@{}}#2\end{tabular}}

\algnewcommand\algorithmicforeach{\textbf{for each}}
\algdef{S}[FOR]{ForEach}[1]{\algorithmicforeach\ #1\ \algorithmicdo}
\algnewcommand{\LeftComment}[1]{\textcolor{gray}{\Statex \ #1}}
\usepackage{amsmath}
\usepackage{subfigure} 
\usepackage{multirow}
\usepackage{booktabs}
\usepackage{url}

\usepackage{caption} 
\begin{document}

\title{Enhancing GUI Exploration Coverage of Android Apps with Deep Link-Integrated Monkey}

\author{Han Hu}
\email{han.hu@monash.edu}
\affiliation{%
  \institution{Monash University}
  \streetaddress{Wellington Road}
  \city{Clayton}
  \state{Victoria}
  \country{Australia}
  \postcode{3800}
}

\author{Han Wang}
\email{han.wang@monash.edu}
\affiliation{%
  \institution{Monash University}
  \streetaddress{Wellington Road}
  \city{Clayton}
  \state{Victoria}
  \country{Australia}
  \postcode{3800}
}

\author{Ruiqi Dong}
\email{rdong@swin.edu.au}
\affiliation{%
  \institution{Swinburne University of Technology}
  \streetaddress{John Street}
  \city{Hawthorn}
  \state{Victoria}
  \country{Australia}
  \postcode{3122}
}

\author{Xiao Chen}
\email{xiao.chen@monash.edu}
\affiliation{%
  \institution{Monash University}
  \streetaddress{Wellington Road}
  \city{Clayton}
  \state{Victoria}
  \country{Australia}
  \postcode{3800}
}

\author{Chunyang Chen}
\email{chunyang.chen@monash.edu}
\affiliation{%
  \institution{Monash University}
  \streetaddress{Wellington Road}
  \city{Clayton}
  \state{Victoria}
  \country{Australia}
  \postcode{3800}
}

\begin{abstract}
Mobile apps are ubiquitous in our daily lives for supporting different tasks such as reading and chatting. Despite the availability of many GUI testing tools, app testers still struggle with low testing code coverage due to tools frequently getting stuck in loops or overlooking activities with concealed entries. This results in a significant amount of testing time being spent on redundant and repetitive exploration of a few GUI pages. 
To address this, we utilize Android's deep links, which assist in triggering Android intents to lead users to specific pages and introduce a deep link-enhanced exploration method. This approach, integrated into the testing tool Monkey, gives rise to Delm (Deep Link-enhanced Monkey).
Delm oversees the dynamic exploration process, guiding the tool out of meaningless testing loops to unexplored GUI pages. We provide a rigorous activity context mock-up approach for triggering existing Android intents to discover more activities with hidden entrances. We conduct experiments to evaluate Delm's effectiveness on activity context mock-up, activity coverage, method coverage, and crash detection. The findings reveal that Delm can mock up more complex activity contexts and significantly outperform state-of-the-art baselines with 27.2\% activity coverage, 21.13\% method coverage, and 23.81\% crash detection.
\end{abstract}


\maketitle

\section{Introduction}
\label{sec:intro}

Mobile applications (apps) have become an integral part of our daily lives and the industry.
However, ensuring their reliability has become a significant challenge. 
Apps are developed rapidly and updated frequently due to the pressure of time-to-market, which leads to a lack of pre-release testing~\cite{planning2002economic}. 
Moreover, Android apps are event-driven systems with complex graphical user interfaces (GUIs) that interact with diverse environments and devices. 
These factors can potentially result in various app flaws, such as unexpected crashes, conflicting UIs, and inappropriate GUI layouts that can negatively impact user experience and lead to substantial user churn~\cite{tian2015characteristics, el2020relationship}. 
Therefore, assuring the quality of Android apps has become increasingly important and urgent.

Android GUI testing, which involves simulating human interactions to detect potential faults in Android apps, has become a vital testing method for developers to ensure the quality of their apps~\cite{androidTesting}. 
When developers test their own apps, one of the key factors determining the effectiveness of Android GUI testing is its code coverage, including activity and method coverage~\cite{wang2018empirical}. 
Studies have shown that the higher the code coverage, the more flaws can be discovered~\cite{codeCoverage, kowalczyk2018configurations, choudhary2015automated}, as no algorithm is capable of detecting vulnerabilities in the code that has not been executed. 
However, the low code coverage of existing Android GUI testing tools, which is always below 20\%~\cite{wang2018empirical}, severely hinders their testing effectiveness, especially testing tools based on dynamic exploration algorithms~\cite{mao2017crowd, mao2016sapienz, azim2016ulink, su2017guided, yan2020multiple}.

\begin{figure}
    \centering
    \includegraphics[width = 0.9\linewidth]{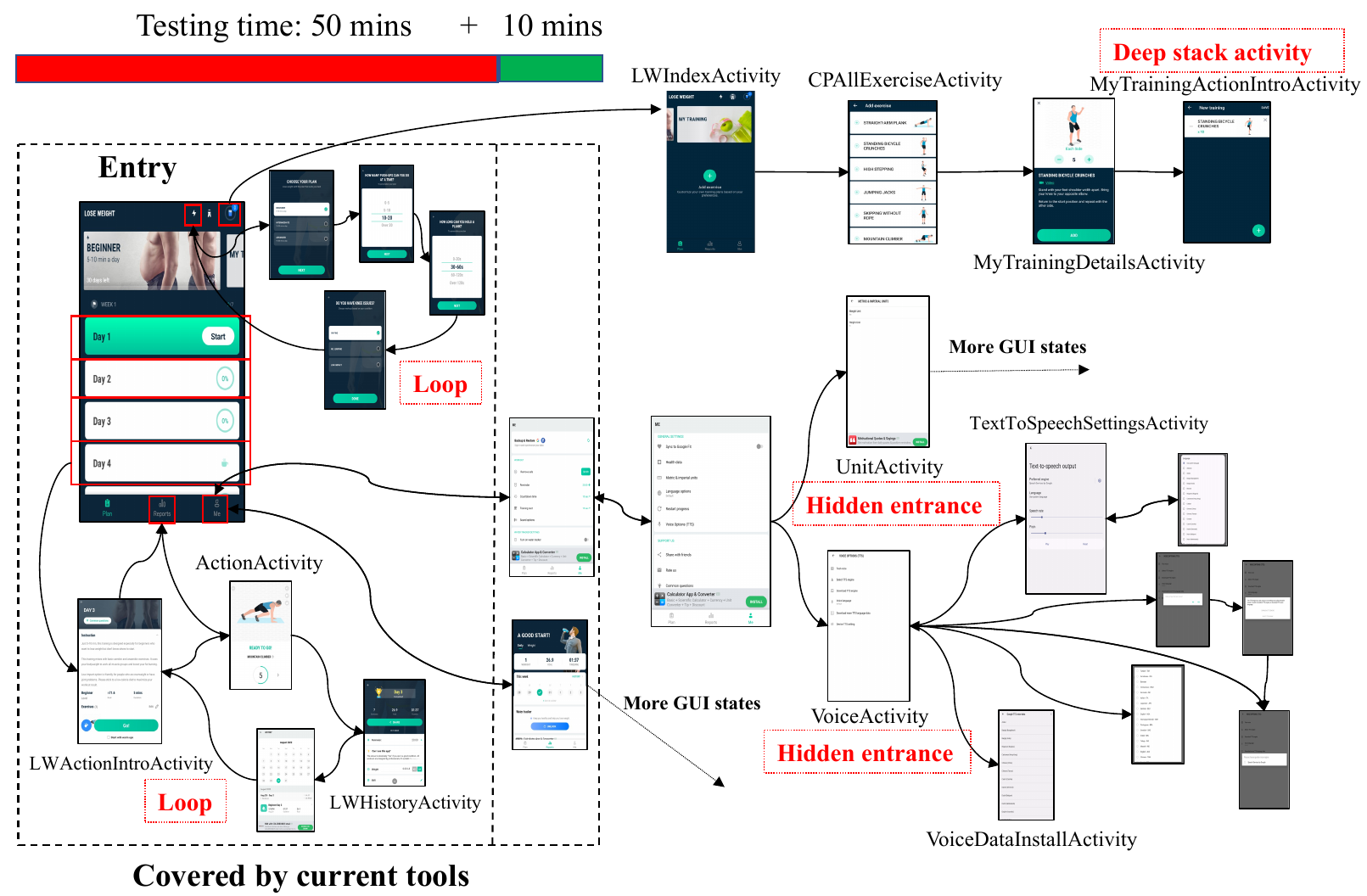}
    \caption{A partial GUI transition graph of the app 'Lose Weight App for Men'. Existing tools always fall into loops between common GUI pages, making it hard to visit some activities with hidden entrances like \emph{VoiceActivity} and some deep stack activities like \emph{MyTrainingActionIntroActivity}, resulting in low code coverage. 
    }
    \label{fig:intro}
\end{figure}

According to our observation, there are three main factors resulting in the low code coverage of the existing testing tools.

First, certain app activities are buried deep within the interface, requiring users to navigate through multiple GUI pages before accessing them. This can make these activities challenging to discover.
For example, accessing the \emph{MyTrainingActionIntroActivity} in  the app ``Lose Weight App for Men''~\cite{loseweight}, as illustrated in Fig.~\ref{fig:intro}, requires the testing tool to navigate through a minimum of three GUI activities. Each activity entrance is intricately positioned within the exponentially expanding investigation branches of the app.

Second, as app GUI pages become more and more complicated,
some activity entries are obscured by complicated UI components, necessitating a series of pre-events (e.g., swipes and taps) to trigger them.
Failing to access these activities can lead to missing all the subsequent GUI pages within the app.
For example, as shown in Fig.~\ref{fig:intro}, testing tools are required to follow a specific sequence: firstly, clicking on the right-most bottom tab to access another fragment, then sliding down the screen and precisely clicking on 'Voice Options (TTs)' and 'Metric \& imperial units' to enter the activities \emph{VoiceActivity} and \emph{UnitActivity}.
Failure to explore \emph{VoiceActivity} will consequently lead to missing all subsequent activities and GUI pages related to \emph{VoiceActivity}.

Third, some current tools have tended to engage in endless cyclical patterns throughout the exploration process~\cite{wang2021vet}. 
In these instances, the tools find themselves in a state of stagnation, circling through a limited subset of GUI pages for an extended testing period, without realizing substantial progress.
In Fig.~\ref{fig:intro}, we demonstrate how one popular GUI testing tool, Monkey~\cite{monkey}, explores the app 'Lose Weight App for Men' for one hour. 
During this time, the main pages and a few surrounding activities are repeatedly explored for approximately 50 minutes, while the testing of other GUI pages and codes is limited to 10 minutes.
App developers have the option to create customized GUI test scripts to improve test code coverage. However, these scripts face challenges when migrating to other apps and are vulnerable to frequent app updates, making them fragile in maintaining reliable testing capabilities.

A deep link is a type of Android intent~\cite{intent} that employs a uniform resource identifier (URI) to directly navigate users from inside or outside of the app to a specific app page~\cite{deeplink}. 
Inspired by this characteristic, we develop a deep link-enhanced GUI testing add-on, which is subsequently integrated into Monkey test~\cite{monkey}, engendering an advanced variant termed Delm to address the aforementioned challenges.
Developers first leverage our add-on to perform a static analysis of the app's source code to gather required information.
During dynamic exploration, we employ a guided exploration algorithm to oversee the entire testing process. If the enhanced testing tool gets stuck in an unproductive loop or overlooks certain Android activities, Delm intervenes temporarily, guiding the tool to unexplored GUI pages. 
Then, it ceases to intervene, allowing the enhanced tool to execute its testing strategy to continue exploring the app.

We recognize that existing Android intents in the app can be triggered not only internally but also externally through the Android Debug Bridge (ADB) tool~\cite{intent, adb}.
To enter an undiscovered activity, Delm follows approaches proposed in related works~\cite{yan2020multiple, yang2018static} to trigger the existing intent receivers from the app outside via associated deep links.
However, it is essential to acknowledge that directing the behavior of an app via external intent triggers may cause additional app crashes, resulting in a few false positives of detected app crashes~\cite{intent, AndroidContext, mockContext}.
To avoid possible false positive crashes, we propose a rigorous approach to check and mock up required activity contexts.
Before triggering intents, Delm verifies whether all activity contexts of intents are prepared.
Delm only allows intents to be triggered where all contexts have been explicitly prepared.




We evaluate the effectiveness of Delm's context mock-up and its ability to avoid false positive crashes using third-party benchmarks, namely IntentBench~\cite{intentBench}, IndustrialApps~\cite{wang2018empirical} and Themis~\cite{su2021benchmarking}.
Our results demonstrate that Delm outperforms existing tools in terms of case-specific activity context mock-up and successfully avoids all false positive crashes while detecting more true positive crashes.
To evaluate the effectiveness of activity coverage in real-world apps, we
perform experiments on close-sourced Google Play~\cite{googleMapURL} industrial apps.
The results show that Delm achieves 27.2\% activity coverage, surpassing the best baseline by 41.37\%. 
To evaluate the effectiveness of method coverage and crash detection, we
conducted tests using Delm on two datasets: the widely-used AndroTest~\cite{wang2018empirical, su2017guided, yan2020multiple} dataset and a set of Google Play industrial apps.
The results indicate that our tool covers 21.13\% methods in the app, outperforming the best baseline by 11.39\%.
Delm also successfully detects 78 identified crashes, demonstrating a 23.81\% improvement over the best baseline, resulting in an increase of 15 crashes detected.

The contribution of this paper are fourfold:
\begin{itemize}
    \item 

    We integrate deep links into Android GUI testing and present a straightforward yet highly effective approach to enhance the code coverage of existing testing tools.
    
    \item 

    We propose a more efficient mock-up method to accommodate various types of Activity contexts. This method not only improves testing effectiveness but also provides valuable insights for future related research.

    \item  

    Through comprehensive experiments on real-world Android apps, we demonstrate the promising performance of our approach. Our results showcase superior testing capabilities when compared to state-of-the-art baselines.

    \item 

    Our tool\footnote{\url{https://github.com/huhanGitHub/guidedExplore}} has been made publicly accessible to stimulate further research, facilitate experiment replication, and serve practical Android app testing scenarios.
    
\end{itemize}

 \section{Background}
\label{sec:background}
This section introduces the background of Android contexts and deep links.

\subsection{Activity Contexts}
Android context, provided by the Android system, serves as an interface to an app's environment~\cite{AndroidContext}.
It enables Android developers to access application-specific resources and perform application-level operations, including launching activities, broadcasting and receiving intents, and more.
In the Android context, four types of contexts have particular significance, and may influence the launch results for a specific activity~\cite{yan2020multiple, AndroidContext}:

\textbf{ICC Message}: The Inter-Component Communication (ICC) message relies on the caller activity and consists of a series of data carried by an Android Intent object~\cite{octeau2013effective}. This facilitates communication between different components within the app, allowing them to exchange data and interact effectively.

\textbf{Device Configuration}: The device configuration context represents the current settings and configurations of the device, such as screen brightness, GPS status, and other device-specific parameters. Understanding the device configuration is crucial for designing adaptive and user-friendly apps.

\textbf{Activity Stack}: The activity stack refers to the sequence of activities that have been visited before the current activity. It is also called Android back stack~\cite{actvityStack}. It plays a pivotal role in managing the navigation flow and the back stack behavior of the app, ensuring a smooth user experience during app usage.
The activity stack in Android functions like a stack of cards, where starting a new activity places it on top of the stack (like placing a card on top of the pile), and finishing an activity or pressing the back button removes the top activity and reveals the previous one, akin to navigating from a product detail page back to the product list in a shopping app.
This structure facilitates a coherent and intuitive navigation experience, enabling users to traverse their activity history in a manner consistent with conventional user interface paradigms.

\textbf{Global Data}: Global data is shared and accessible throughout the app's lifecycle. 
It includes important information such as the user's login status, preferences, and other app-wide variables that need to be maintained consistently across various activities.

In this paper, our efforts are concentrated on mocking up the ICC messages contained within intent objects, avoiding the mock-up of global data and the activity stack. Instead, we design the guided exploration algorithms to prevent these elements from being inadvertently engaged by deep links.



\subsection{Deep Link}
\label{sec:deepLink}
The Android intent is an abstract description of an operation that enables communication between different components within an Android app. It serves as a mechanism to request actions from other components~\cite{intent}.
Deep links are a specific subset of activity-triggering intents, designed to enable direct access to particular content or pages within an app from external sources a URI (Uniform Resource Identifier)~\cite{deeplink}. 
This functionality stands in contrast to regular intents, which are broadly used within Android apps for various types of activity navigation and inter-component communication. 
Deep links are particularly useful for navigating users directly to specific locations within an app from web pages, emails, social media, or other external apps, enhancing the user experience by providing a seamless transition. In the context of our research, we leverage deep links as a method to externally trigger certain Android activities for GUI testing purposes. Our approach, being a third-party solution, does not integrate directly into the native app codebase. Instead, it utilizes deep links to simulate user interactions with the app from outside, a method constrained by Android's architectural and security frameworks. 
In Android, deep links allow the developers to define intent filters to declare the available links to their app. 
Developers can also define data types inside the intent filter as parameters for deep links.



 \section{Approach}


\begin{figure}
    \centering
    \includegraphics[width=0.9\linewidth]{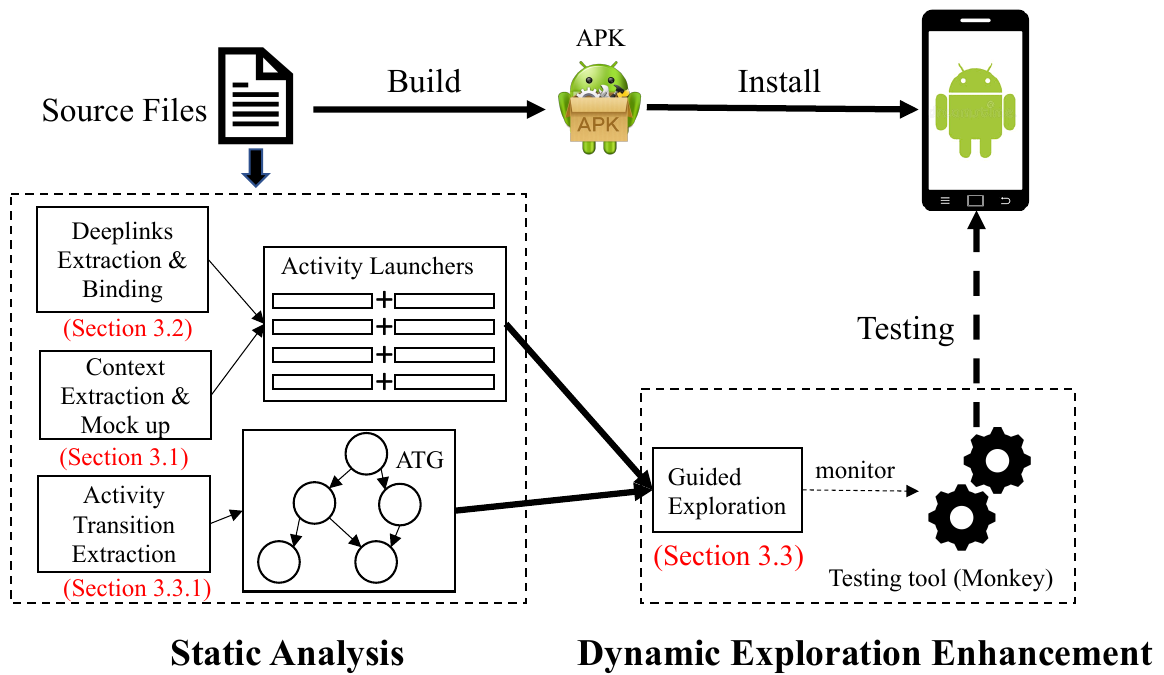}
    \caption{The workflow of Delm}
    \label{fig:framework}
\end{figure}

Fig.~\ref{fig:framework} depicts the workflow of Delm, which comprises two phases: \emph{Static Analysis} and \emph{Dynamic Exploration Enhancement}. 
Delm is principally engineered to accommodate the source code of apps as input. However, it also possesses the capability to operate on a selection of injectable APKs.
When dealing with these APKs, our initial step is to decompile them into human-readable smali files which can elucidate the runtime execution of an Android app~\cite{hoffmann2013slicing}.
Owing to the smali code's interpretability, similar to related works~\cite{yan2020multiple, chen2019storydroid, chen2021accessible}, we extend the static analysis performed on Java code to the smali code for extracting the required contextual information.


In the following sections, we first illustrate how we verify and mock up the required activity contexts in apps in Section~\ref{sec:context}.
Then, we describe the process of extracting and binding deep links in existing Android intents in Section~\ref{sec: deeplinkExtra}.
Finally, we explain our dynamic exploration enhancement algorithm in Section~\ref{sec:guidedExplo}.




\subsection{Activity Context Analysis and Mock-up}
\label{sec:context}

When Delm triggers an intent to launch an activity from outside, it should provide appropriate activity contexts to prevent launch failures and false positive crashes~\cite{AndroidContext}. 
Consequently, we introduce four types of context in Section~\ref{sec:background}: ICC message, device configuration, activity stack, and global data that may influence activity launch. 

To safeguard the coherence of the activity stack and global data with the dynamically launched activity, we adhere strictly to the path delineated in the app's predefined Activity Transition Graph (ATG). Such adherence guides the selection of subsequent activities primed for launching. When Delm intervenes during GUI exploration, it references the previously extracted ATG and rigidly follows the app's predefined ATG to identify the next proximate and accessible activity.
In the event that all appropriate adjacent activities become unidentifiable, Delm proceeds to an early conclusion of the GUI exploration phase.
Regarding device configuration, we follow the instrumented-based testing tool Fax~\cite{yan2020multiple} to grant the app all required permissions during the normal dynamic exploration phase.

\begin{figure}
    \centering
\includegraphics[width = 0.9\linewidth]{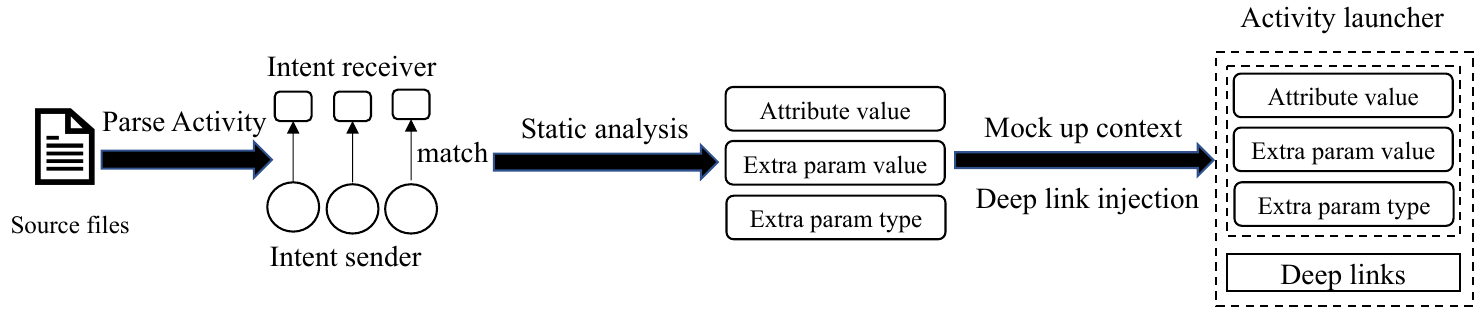}
    \caption{The pipeline of ICC message context checking and mock-ups}
    \label{fig:contextMock}
\end{figure}

Android inter-component communication (ICC) plays a vital role in launching activities, and incorrect ICC messages are the primary cause of launch problems~\cite{yan2020multiple, octeau2013effective}. Any inconsistency in the type or amount of data required by the target app page during ICC may result in launch failure or false positive crashes~\cite{mockContext}. 
Fig.~\ref{fig:contextMock} illustrates the pipeline for mocking up and checking the local activity context of ICC messages and binding deep links. 
The methodology begins by analyzing intent senders and receivers in the source code to identify essential contexts. We then create potential launchers for each activity, combining mock contexts and deep links. These custom launchers are prepared to initiate activities for exploration in scenarios where dynamic reachability is limited.
In Section~\ref{sec: deeplinkExtra}, we elucidate the methodology employed for the extraction and binding of deep links to pre-existing intents, purposed for external launching. To obviate false positive crashes, our approach adopts a conservative stance, eschewing the construction of contexts for implicit intents with wildcards and particular intents that suffer from indistinct attributes or absent extra parameter values.
However, it is noteworthy that Delm retains the capacity to fabricate contexts and pinpoint suitable activities by analyzing the wildcard expressions of implicit intents. 
This careful handling ensures that the testing process remains reliable and minimizes the risk of encountering false positive crashes.

The context data carried by the intent could be divided into two types: attributes and extra parameters~\cite{intent}.

\textbf{Attributes}
The attributes encompass \emph{action}, \emph{category}, \emph{data}, and \emph{type}~\cite{AndroidContext}. 
When processing implicit intents without wildcards in Android manifest files, we extract verification details from each activity's \emph{intent-filter} tag, which include \emph{action}, \emph{category}, and \emph{data} attributes. 
This serves to ascertain requisite properties for intent receivers. 
Leveraging this information as a criterion, we scan through all the activities in the app, identifying the intent senders that satisfy these conditions.
Regarding explicit intents within the source code of Android activities, we rely on the initialization constructor of the intents, coupled with \emph{setComponents}, \emph{setClass}, and \emph{setClassName} functions, to designate the corresponding receivers and senders of these intents.

To trace and verify the attribute values that an intent receiver demands, we construct a definition-use (def-use) chain~\cite{milanova2005parameterized} within the coupled sender activity. As a standard technique in dataflow analysis, this chain helps us track how values assigned to intent attributes evolve throughout the app's execution flow.

We employ custom scripts to parse the app’s source code or decompiled smali codes, identifying the declarations and invocations of critical attribute-setting methods, including \emph{setAction}, \emph{setType}, \emph{setData}, \emph{setFlags}, and \emph{setIdentifier}.
As per the Android developer documentation~\cite{intent}, the second parameters/registers of these methods hold values depicting the ICC information, which is destined for intent receivers. 
Our analysis progresses by tracking the alterations in these parameters/registers within both Java and smali files before their final use in attribute-setting methods.
This tracking begins from the moment parameters are initialized or assigned to a certain constant value within the sender activity, following their journey across distinct code blocks, loops, conditional statements, and method invocations.
As parameters are manipulated, combined, or altered, we record each transformation, capturing the lineage of data changes up to the point where these values are finally employed in configuring intent attributes. 
By understanding how and when parameter values are set, modified, and ultimately utilized, we can accurately determine the final attribute values intended for the intent receiver, ensuring a thorough and precise mapping of the ICC information within the app's workflow.



\textbf{Extra Parameters}
In Android, an extra parameter represents an auxiliary information bundle, articulated in a key-value format~\cite{AndroidContext}. 
To launch an activity associated with an intent, it is necessary to employ both the accurate key and corresponding value within the extra parameter.
Extra parameters can be subdivided into three categories: primary, object and bundle.
The primary type is characterized by a key-value pair, wherein the key is a \emph{String} type, and the value could be any Java primitive or non-primitive type, such as \emph{int}, \emph{String}, \emph{ArrayList}, and the like.
The object type corresponds to a key-value pair, where the key is classified as a \emph{String} type, and the value could be either a Serializable/Parcelable class object or an array of such objects.
Contrastingly, the bundle type represents a compendium of key-value pairs, wherein each pair can encapsulate a primary type, bundle type, or object type~\cite{bundle}. 

To pinpoint the value and type of extra parameters, we adopt a methodology that is the same as the attribute identification process. We initially locate the extra parameter setting methods in the source code, which allows us to obtain the necessary extra parameters, followed by tracking the alterations of these variables within the sender activities.
In relation to object parameters, we trace the source code pathway of these objects and leverage Java reflection techniques~\cite{JavaReflection} to construct these objects, thereby enabling us to emulate object ICC messages.

In the activity context analysis, the management of uncertain context values, notably implicit intent filters, introduces significant risk. The process of parsing <action> and <data> tags from implicit intents to infer required data formats, while constructive, is hampered by challenges in accurately determining data values. This limitation can substantially increase the likelihood of generating false positives. Successfully activating an implicit intent with a substitute mocked-up value and initiating the related activity might unintentionally escalate the risk of false positive crashes.
This, in turn, exacerbates the manual verification efforts required. Consequently, a more reserved strategy is employed in the face of uncertain context mocking. Particularly in cases involving dynamic data, our approach is to ascertain at least one definitive value for intent mocking. In instances where such precision is unattainable, the decision is made to abstain from the mock-up of those intents.




\begin{figure}[htbp]
    \centering
    \includegraphics[width=0.9\textwidth]{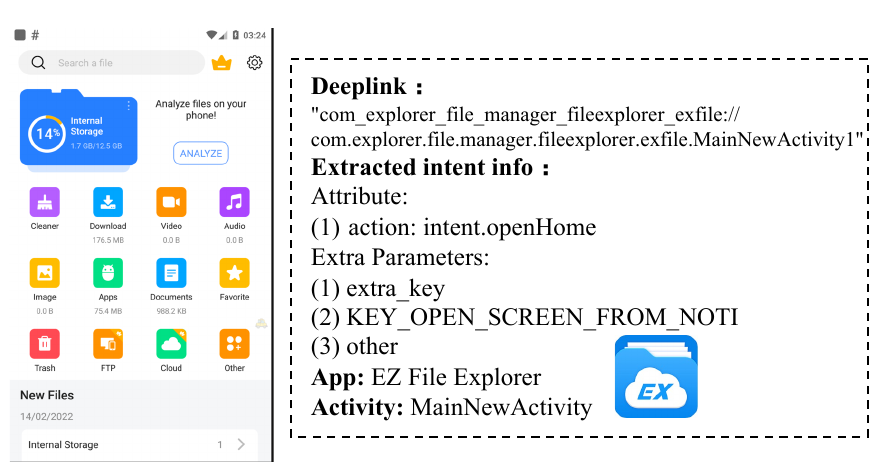}
    \caption{An example of deep link and extracted intent context information in \emph{MainActivity} of app `EZ File Explorer'.}
    \label{fig:activityLauncher}
\end{figure}

Fig.~\ref{fig:activityLauncher} depicts a successful launch of the activity \emph{MainNewactivity} in the app `EZ File Explorer'~\cite{ezfile} by Delm. 
The triggered deep link for this activity is a unique URL defined by our tool.
The context attribute \emph{intent.openHome}, defined by app developers, is set as the \emph{action} attribute of the launch intent.
We also extract all user-defined extra parameters (\emph{extra\_key}, \emph{KEY\_OPEN\_SCREEN\_-FROM\_NOTI} and \emph{other}) in this activity.


\subsection{Deep Link Extraction and Binding} 
\label{sec: deeplinkExtra}
Based on the analysis of specific identifiers related to deep links in Android Manifest files, we extract the available deep links. Associating a suitable deep link with each intent can increase the success rate of external intent triggering~\cite{deeplink}. Thus, we automatically integrate a deep link for each Android intent to facilitate external intent triggering during the enhanced exploration.


\subsubsection{Deep Link Extraction}
Informed by preceding research and the Android documentation~\cite{ma2018aladdin, deeplink, intent}, Android intents, which facilitate deep links, necessitate certain elements and attributes to be included in the Manifest file. These requisite elements are encapsulated within the \emph{intent-filter} tag and must encompass the \emph{android.intent.action.VIEW} action as well as the \emph{android.intent.category.-BROWSABLE} category. 
Additionally, deep links necessitate the \emph{data} tag to contain \emph{@android:scheme} and \emph{@android:host} attributes. Consequently, we instigate a comprehensive analysis of the entire Manifest file, from which we extract all instances of \emph{activity} tags that adhere to these defined patterns.

\subsubsection{Deep Link Binding}

Not all activities within an app are initially set up with deep links. In instances where an activity lacks a pre-configured deep link, our methodology involves the binding of a new deep link to these activities.
First, we begin by retrieving applicable activities that have been discovered within the Manifest file and subsequently, for each of these activities, we modify the \emph{@android:exported} attribute to \emph{True}. 
This allows the activity to be accessible from outside the application.
Next, we insert an \emph{intent-filter} tag inside each \textit{activity} tag. Within this \emph{intent-filter} tag, we include a \textit{data} tag with \emph{@android:scheme} and \emph{@android:host} attributes specified for the injected deep links.
Additionally, we ensure that the \emph{intent-filter} includes the required \emph{action} and \emph{category} attributes, as necessary for defining the deep links.

\begin{figure}[htbp]
    \centering
    \includegraphics[width=0.8\textwidth]{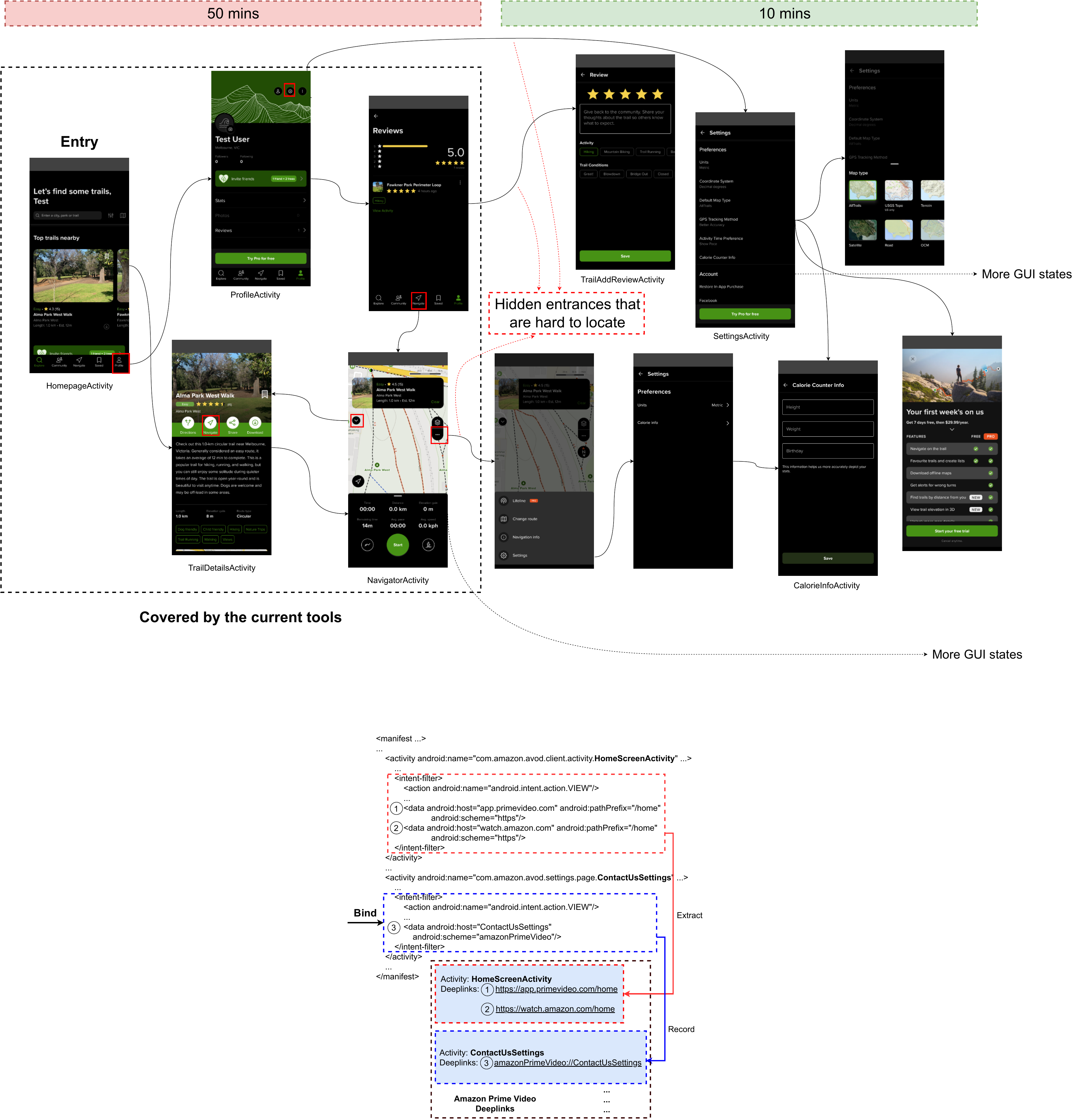}
    \caption{An example of how Delm extracts existing deep links \textcircled{1} and \textcircled{2} and binds new deep link \textcircled{3} in the Manifest file}.
    \label{fig:deep linkExample}
\end{figure}

Fig.~\ref{fig:deep linkExample} presents a process of deep link extraction and binding in the app `Amazon Prime Video'. 
Developers of this app declare two deep links for the \emph{HomeScreenActivity}. 
Accessing either of the links shall direct users to the \emph{HomeScreenActivity} on an Android Device.
For the \emph{HomeScreenActivity}, we extract the existing deep links defined in the original code (\emph{deep links 1, 2}). 
For \emph{ContactUsSettings}, our tool detects existing intents pointing to this activity but no related deep links. 
Therefore, we inject and bind an \emph{intent-filter} inside the activity tag for these discovered intents (\emph{deep link 3}). 
We expose activities to the outside of the app with such an approach. 

\subsubsection{Contexts Setup and Activity Launch Process}
In this section, we delineate the whole process involved in configuring the context and launching activities within an Android app through deep links, structured into four essential phases.
(1) Context Analysis: The foundation of our approach involves a detailed examination of the app's source or its decompiled smali code. This step is critical for gathering the necessary context information required for launching Android activities effectively.
(2) Intent Filters Identification: Initially, we analyze the app's \emph{AndroidManifest.xml} to identify existing deep links associated with activities. For activities lacking deep links, we dynamically declare intent filters, detailing the actions, categories, and data the activity is equipped to handle.
(3) Data Transmission: URIs within deep links serve to transmit data, enabling targeted activities to configure their context accordingly. For example, the deep link `http://www.example.com/detail?id=123` directly initiates `DetailActivity` with an `id` parameter. For pre-existing deep links in the app's manifest, we bind the necessary data within their \emph{<data>} elements to facilitate data transmission.
(4) Intent Handling by Activities: When triggered by a deep link, an activity will receive an intent that carries data. This data can then be parsed to configure the activity's contexts. Consequently, the activity associated with the deep link is launched, with its contexts initialized.

\subsection{Dynamic Exploration Enhancement}
\label{sec:guidedExplo}

In this section, we detail the methodology employed to enhance testing tools in dynamic exploration, aiming to increase code coverage. Initially, we analyze the Application Transition Graph (ATG) of apps to preserve the global activity contexts, encompassing both the activity stack and global data. We then introduce a novel approach to identifying unique Android GUI states to prevent unnecessary repetition in exploration. Additionally, we describe our strategy for determining when a testing tool becomes trapped in a loop during exploration. Finally, we explain how Delm actively intervenes in the exploration process to improve its efficiency.

\subsubsection{Activity Transition Graph}
Following the related work StoryDroid~\cite{chen2019storydroid}, we utilize IC3~\cite{ic3} and Soot~\cite{Soot} to extract the app's ATG. 
This strategic approach assists us in identifying the specific activity to launch via deep links, predicated on its correlation with the activity presently being tested. 
It is noteworthy to mention that ATGs extracted through static analysis might be incomplete. Thus, we also supplement our ATG with reliance on the runtime activity stack during the dynamic exploration of the app. In light of the characteristics inherent to the activity stack, it follows that two activities positioned adjacently within the stack should likewise be represented as adjacent within the ATG. This strategy exploits the inherent relationship between the activity stack and navigation, dynamically refining the ATG to provide a more exhaustive depiction of potential activity transitions not discernible through static analysis alone.

\subsubsection{Unique GUI state}
Following prior works~\cite{machiry2013dynodroid, pan2020reinforcement, wang2021vet}, 
we define each GUI page on Android devices, consisting of its GUI metadata hierarchy tree and component attributes, representing one GUI state.
The uniqueness of a GUI state is ascertained by hashing the structure of its GUI hierarchy tree. It is important to note that within a single activity, alterations in the structure of the current GUI page will instigate the generation of a new unique state~\cite{dong2020time}.


\subsubsection{Loop Detection}

\begin{algorithm}[h]
\caption{Dynamic Loop Detection Algorithm}
\label{alg:loopDetection}
\begin{algorithmic}[1]
\Require currentActivity ($currentActivity$), enhancedTestingTool ($testingTool$)
\Require activityTransitionGraph ($activityTransitionGraph$), timeoutThreshold ($timeoutThreshold$)
\Require visitedActivityStack ($visitedActivityStack$)
\LeftComment{Predefined constants: maximum repetition threshold and maximum queue size}
\State $maximumRepetitionThreshold \gets \alpha$, $maximumQueueSize \gets \beta$
\State $stateQueue \gets \{\}$, $currentState \gets \Call{Dump}{currentActivity}$

\While{$event \gets testingTool.\Call{GenerateEvents}{timeoutThreshold}$}
    \State $previousState \gets currentState$
    \State $stateQueue.\Call{Append}{previousState}$    
    \State $currentState \gets testingTool.\Call{ExecuteEvent}{event}$
    \LeftComment{Update ATG and activity stack on activity change}
    \If{$currentState.\text{ACT} \neq previousState.\text{ACT}$}
        \State $activityTransitionGraph.\Call{Update}{previousState, currentState}$
        \State $visitedActivityStack.\Call{Add}{previousState.\text{ACT}}$
    \EndIf
    \LeftComment{Detect and handle state loops}
    \If{$\Call{Size}{stateQueue} > maximumQueueSize$}
        \State $numRepeated \gets stateQueue.\Call{MaxRepeatedState}{}$
        \If{$numRepeated > maximumRepetitionThreshold$}
            \State $stateQueue.\Call{Clear}{}$  
            \State \Return True
        \Else
            \State \Return False
        \EndIf
    \EndIf
\EndWhile
\end{algorithmic}
\end{algorithm}

Algorithm~\ref{alg:loopDetection} demonstrates how to dynamically detect whether the tool gets stuck in a loop. 
We will force intervention in the tool's exploration process when loops or timeouts occur, reducing useless repeated exploration and time wasting.
The algorithm operates with several essential inputs, including the currently focused activity ($currentActivity$), an advanced testing tool ($testingTool$), the Activity Transition Graph ($activityTransitionGraph$), a specified timeout duration ($timeoutThreshold$), and a collection of visited activity stack during the exploration process ($VisitedActivityStack$). 
Key to the loop detection strategy are two empirically determined constants: the maximum repetition threshold ($maximumRepetitionThreshold$) and the maximum queue size ($maximumQueueSize$). 
We consider a loop to occur when a single state or a fixed set of GUI states is explored repeatedly beyond the maximum repeated threshold. The specific values chosen for these constants are carefully calibrated to balance the trade-offs between accuracy and efficiency in the testing process. Default values of maximum repetition threshold and maximum queue size are empirically set to 50 and 200.

When the tool begins to execute testing events on the app, Delm monitors the explored GUI states and appends these states to a state queue ($stateQueue$) (lines 3-6).
Subsequent to each event execution, if the current GUI state changes to new activities, Delm collects new visited activity in the activity stack ($visitedActivityStack$) for subsequent possible intervention (lines 7-10). 
Whenever a novel state is appended to the state queue, Delm conducts an assessment to verify whether the quantity of states in the queue surpasses the predefined upper limit ($maximumQueueSize$) (lines 11-12). 
Upon surpassing this threshold, and if a state within the queue has been reiterated beyond the maximum repeated threshold ($maximumRepetitionThreshold$), it is inferred that the tool is caught in a loop. In such instances, an intervention is triggered to pause the exploration (lines 13-17).

\begin{algorithm}[h]
\caption{Guided Exploration Algorithm}
\label{alg:guidedExploration}
\begin{algorithmic}[1]
\Require activityTransitionGraph ($activityTransitionGraph$), androidApplication ($app$)
\Require enhancedGUITestingTool ($testingTool$), timeoutThreshold ($timeoutThreshold$)
\Require mockUpActivityContextData ($mockUpData$)

\State $visitedActivitiesStack \gets \{\}$
\State $entryActivity \gets$ \Call{Launch}{$app$}
\LeftComment{Dynamically launch the entry activity at the start}
\State $visitedActivitiesStack$.Add($entryActivity$)
\State $currentActivity \gets entryActivity$ 

\While{\textbf{not} $timeoutThreshold$}
    \State $loopDetected \gets$ \Call{LoopDetection}{$currentActivity$, $testingTool$, $activityTransitionGraph$, $timeoutThreshold$, $visitedActivitiesStack$}
    
    \If{$loopDetected$}
        \LeftComment{Find an adjacent activity in ATG and its intent data}
        \State $nextActivity \gets$ \Call{FindAdjacentActivity}{$activityTransitionGraph$, $mockUpData$}
        \If{$nextActivity$ in $visitedActivitiesStack$}
\LeftComment{if the selected next activity has been visited and is in the activity stack, pop this activity from the activity stack and launch it dynamically}
            \State \Call{DynamicLaunch}{$nextActivity$}
        \Else
            \LeftComment{Use the corresponding deep link to launch the activity if not previously visited}
            \State \Call{DeeplinkLaunch}{$nextActivity$}
        \EndIf
        \State $currentActivity \gets nextActivity$
    \EndIf
\EndWhile
\end{algorithmic}
\end{algorithm}

 \subsubsection{Guided Exploration}


When the testing tool is stuck in an exploration loop, Delm is activated to guide exiting the loop. 
The objective of incorporating guided exploration within our methodology serves two primary purposes. Firstly, it complements activity mock-up and loop detection to enhance the efficiency of app GUI exploration, enabling the discovery of a wider array of activities. 
Secondly, it ensures the exploration process strictly follows the app's inherent activity transition graph without altering the app's original architecture. 
This algorithm is pivotal in maintaining the app's designed behavior, particularly in relation to global data and the activity stack, thereby minimizing the likelihood of introducing false positives through unintended interference.

The process of guided exploration is outlined in Algorithm~\ref{alg:guidedExploration}. The inputs for this algorithm are the activity transition graph ($activityTransitionGraph$), the Android app ($app$), the testing tool ($testingTool$), a timeout threshold ($timeoutThreshold$), and mock-up activity context data ($mockUpData$).
First, we collect dynamic visited activities in the activity stack ($visitedActivityStack$) (line 1).
Delm launches the app dynamically and obtains the entry activity (lines 2-4).
Within the required testing time, we allow the enhanced testing tool to test the app.
During dynamic exploration, Delm constantly monitors the tool to ensure that it does not get stuck in a loop, as illustrated in Algorithm~\ref{alg:guidedExploration} (lines 5-6). 
If the tool becomes stuck in a loop, Delm selects the next accessible adjacent activity $nextActivity$ and its inside intents by referring to the ATG and prepared context data (lines 7-8).
Delm prefers to direct the tool towards the next activity with discovered dynamic transfer paths in the ATG and outside the current loop.
If the selected activity has been visited and is in the activity stack ($visitedActivityStack$), Delm pops the activity in the stack and dynamically launches it from the activity stack (lines 9-10).
It is important to recognize that revisiting an activity previously encountered is not devoid of value. This is because even visited activities can present unexplored states or lead to subsequent activities that have yet to be accessed, offering potential new paths and insights into the app's testing.
If the tool does not visit the activity, Delm uses pre-discovered intents and mock-up contexts to launch it (lines 11-13).
If the app becomes unable to move forward or backward, Delm restarts the app and dynamically navigates to an unexplored activity.
After accessing the next activity, the tool continues exploring the launched activity until it gets trapped in another exploration loop or the test period has expired (lines 14-16).

\begin{figure} 
    \centering
    \includegraphics[width = 0.8\textwidth]{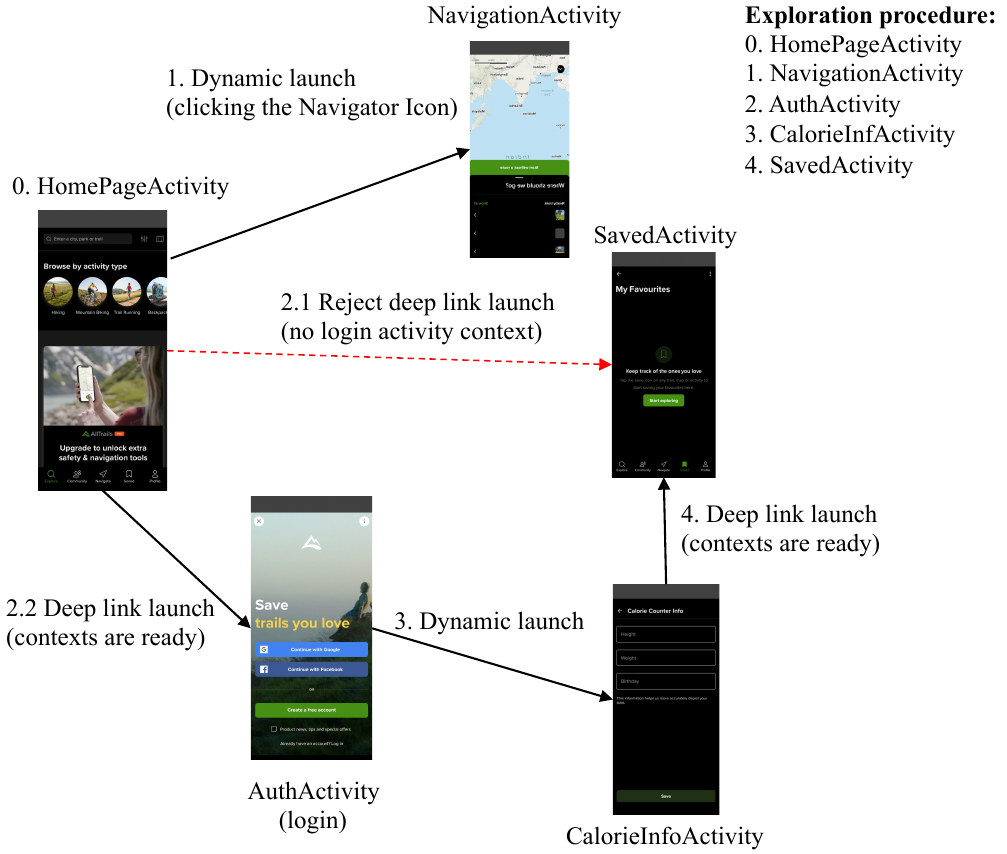}
    \caption{A concrete example of the enhancement exploration process in the app `Alltrails'.
    }
    \label{fig:internalExploration}
\end{figure}

Fig.~\ref{fig:internalExploration} shows a concrete example of the app Alltrails'~\cite{alltrails} guided exploration procedure.
After exploring the entry activity $HomePageActivity$, the testing tool becomes stuck in a loop. 
Delm dynamically guides the tool to the next accessible adjacent activity $NavigationActivity$ via the activity stack. 
However, after a while, the exploration returns to a loop between $HomePageActivity$ and $NavigationActivity$. 
Delm identifies that $HomePageActivity$ has two other adjacent activities, $SavedActivity$, and $AuthActivity$. 
No dynamic entries are discovered for both activities at that moment. 
After checking the context requirements for both activities, it is found that the $SavedActivity$ requires a login before being launched, which is not ready at this stage. 
Therefore, Delm selects the $AuthActivity$ as the next target activity for the tool and launches it with binding deep links to explore. 
After logging in and exploring the $AuthActivity$, the tool discovers a new activity, $CalorieInfoActivity$, and the login context is now ready for the $SavedActivity$. Therefore, the $SavedActivity$ is selected as the next launch candidate for exploration. Following this process, Delm continues to accumulate new access paths for deeper activities in the app.

\section{Evaluation}
In this section, we describe the setup and results of our experiments.
To quantitatively and qualitatively evaluate our tool, we propose three research questions:
\begin{itemize}
    
    \item RQ1: How effective is our approach in activity context mocking up?
    
    \item RQ2: How effective is our approach for activity coverage in real-world apps? 
    
    \item RQ3: How effective is our tool for code exploration and crash detection in real-world apps? 
    
\end{itemize}
In RQ1, we first evaluate our tool's effectiveness in context mock-up.
Second, we check the likelihood of false positives in the crash detection tasks of Delm and the selected baselines. 
In RQ2, we evaluate the effectiveness of activity coverage of our tool in real-world apps. 
The more activities that can be covered, the more GUI components and app features can be tested.
In RQ3, we evaluate the method coverage and crash detection of Delm and how it compares to existing state-of-the-art GUI testing tools.

\subsection{RQ1 Effectiveness of the Context Mock-up}
\label{sec:rq1}


\subsubsection{Benchmark}

Three benchmarks, IntentBench~\cite{intentBench}, IndustialApps~\cite{wang2018empirical} and Themis~\cite{su2021benchmarking} are used to evaluate the effectiveness of the context mock-up of Delm.

To evaluate the ability to mock up contexts, we reuse and expand the third-party benchmark IntentBench, which has been widely adopted in previous works~\cite{yan2020multiple, guo2020improving}.
We also go through all the ICC message-related APIs in Android 10~\cite{Android10} to expand more uncovered context categories in the current benchmark.
This benchmark designs different value check \emph{if-else} blocks with different conditions in each activity, such as getting extra parameters of intent in one loop, getting user-defined attributes, and extra parameters.
Each value check block is regarded as a testing case.
Only if the tool extracts and passes the correct values to activities will we pass these cases, which could be verified by the console outputs of the benchmark app.
For example, as demonstrated in Listing 1, if Case 7 ($(7)\ check\ the\ global\ data$) is mocked up successfully, the console will display 'pass 7'.

The benchmark comprises 8 categories of activity intents: \emph{Attribute}, \emph{Primary extra param}, \emph{Object extra param}, \emph{Bundle extra param}, \emph{Basic + Extra}, \emph{Activity stack}, \emph{Global data}, and \emph{Device configuration}.
Listing 1 provides examples of test cases in each category.
For test cases in the categories \emph{Attribute} and \emph{Primary/Object/Bundle extra params}, tools must extract and mock up the same attribute values for receiving intents to pass these cases.
For test cases in \emph{Activity stack}, tools must ensure the same activity transition path to extract correct global data values.
As shown in the sixth case in Listing 1, to guarantee that the proper value is sent to the current activity, the tool must explore along the correct activity stack (Activity $A \rightarrow B \rightarrow C$).
For test cases in \emph{Global data} and \emph{Device configuration}, tools are required to ensure correct global data and configuration in the current activity to pass test cases.

\lstset{
  basicstyle=\fontsize{8}{8}\selectfont\ttfamily
}
\begin{lstlisting}[language=Java, 
caption={Code examples of test cases in the context mock-up evaluation}, 
captionpos=b,
]
    // (1) check the attribute
    String action = getIntent().getAction();
	if(action.equals("action1"))
			System.out.println("pass 1");
	// (2) check the primary extra parameters
    String str = getIntent().getStringExtra("str");
    int i= getIntent().getIntExtra("i",0);
	if(str.equals("action1") && i ==1)
		System.out.println("pass 2");
	// (3) check the object extra parameters
	// "Person" is a serializable class
    Person p = (Person) getIntent().getSerializableExtra("person");
	if(p.getName().equals("name"))
		System.out.println("pass 3");
	// (4) check the bundle extra parameters
	// "bundle1" is a key-value pair in bundle
    String str = getIntent().getStringExtra("bundle1");
	if(str.equals("bundle1"))
		System.out.println("pass 4");
    // (5) check the attribute and extra
    String action = getIntent().getAction();
    String data = getIntent().getData()
	if(action.equals("action1") && data.equals("uri1"))
			System.out.println("pass 5");	
    // (6) check the activity stack: activity A --> B --> C
    // "Student" is a static class
    // in activity A
    Student.count = 0 
    // in activity B
    Student.count += 1
     // in activity C
     // check the count of class Student
	if(Student.count == 1)
			System.out.println("pass 6");
    // (7) check the global data
     String major = Student.major
     if(major.equals("CS"))
	    System.out.println("pass 7");
	// (8) check the device configuration of GPS 
    LocationManager l = getSystemService(Context.LOCATION_SERVICE)
    System.out.println("pass 8");
\end{lstlisting}

The IndustialApps dataset~\cite{wang2018empirical} encompasses 68 top-rated apps from each category on Google Play. To massively assess the efficacy of activity mock-ups within real-world apps, we extract obtainable intents from these apps. 
It is essential to note that all included apps are updated versions, with updates being current up to February 2023.

To evaluate whether our tool produces false positive crashes, we select Themis~\cite{su2021benchmarking} as the benchmark, which comprises 20 open source apps with confirmed crash bugs and have been widely adopted by related works~\cite{su2021fully, lv2022fastbot2, wang2022detecting}.
Themis contains 52 reproducible crash bugs. 
All these bugs are labeled by the app developers as "critical bugs" (i.e., important bugs), which affect the major app functionalities and a larger percentage of app users.


\subsubsection{Baselines}
Existing context mock-up tools StoryDroid~\cite{chen2019storydroid, chen2021accessible} and Fax~\cite{yan2020multiple} are selected as the baselines in the context mock-up experiment.
StoryDroid employs static analysis to construct the activity transition graph, launch activities, and generate the storyboard for Android apps.
Fax is a multiple-entry testing tool for Android apps by constructing activity launching contexts.
Fax analyzes the activity contexts from the Android Java source code and tests all the activities in the app according to the order of the activity weights calculated.
The functionalities of these baselines correspond to the activity launch mechanisms we investigated.

To evaluate the contribution of our guided exploration algorithm in preserving the activity stack and global data contexts, we introduce a comparative baseline, Random Delm. This variation disables the guided exploration feature, adopting a random approach to activity launch. This comparative analysis aims to highlight the significance of guided exploration in enhancing GUI testing through the effective preservation of global data and activity stack.

False positives in app crash detection may result from any test technique that does not adhere to the native app design, such as instrumentation, fuzzing, etc~\cite{alegroth2015visual}.
Both Fax and StoryDroid have the potential to change apps' behaviors.
Given that StoryDroid is not a testing tool, we only select Fax as the baseline in the false positive experiment.

\subsubsection{Procedure}
For the evaluation of context mock-up, we utilize tools to directly launch the activities of the benchmark app, subsequently conducting a manual inspection to ascertain the count of successful mock-up instances. 
Regarding the massive context mock-up evaluation within the industrial apps, we initially extract all intents present within the apps and determine the required contexts to launch the corresponding activity via smali files. 
Subsequently, tools are employed to sequentially trigger these intents. 
We automatically verify whether the launch is successful through the output of ADB and the changes in the GUI interface.
It's worth noting that in scenarios where the corresponding activities' contexts cannot be accurately mocked up, such cases would be classified as launch failures, despite the actual launching of these activities.

In the false positive evaluation, by referring to the benchmark's documentation, we run tools to test each APK in the benchmark independently for three hours and manually reviewed the results report.
The crash is regarded as a true positive when the tool successfully triggers pre-tagged crash bugs. 
Unknown crashes that cannot be manually reproduced are considered false positives.

\begin{table}[]
\centering
\caption{Results of Mocking Up Activity Contexts. The numbers represent the count of successful mock-up instances, defined as cases where the tool accurately replicates the intended activity context without errors.}
\small
\resizebox{0.7\textwidth}{!}{

\begin{tabular}{|c|c|c|c|c|c|}
\toprule
Category & \#Testing cases & \#StoryDroid & \#Fax & \#Random Delm & \#Delm\\

\midrule
Attribute   & 13            & 10       & 12 & 13 & 13\\
\rowcolor{gray!25} Primary extra param   & 28        & 24      & 27 & 28 & 28\\
Object extra param   & 8           & 0           & 0 & 1 & 1\\
\rowcolor{gray!25} Bundle extra param     & 12       & 11      & 12 & 12 & 12\\
Basic + Extra         & 11         & 7         &  10    & 11 & 11\\
\rowcolor{gray!25} Activity stack     & 2        & 1       & 0  & 0 & 2\\
Global data      & 5      & 3       &  3 & 3 & 5\\

\rowcolor{gray!25} Device configuration     & 2       & 2       &  2 & 2 & 2\\
\midrule 
IntentBench cases     & 81           & 58  & 67 & 70 & 74\\

\midrule 
\rowcolor{gray!25}Real-world cases & 12,070 & 1,935 & 2,072 & 2,171 & 3,003 \\

\bottomrule
\end{tabular}
}
\label{tab:rq1}
\end{table}

\subsubsection{Results of Contexts Mock-up}
As shown in Table~\ref{tab:rq1}, we use the column \emph{Category} to represent the category of testing cases.
The column \emph{\#Testing Cases} indicates the number of testing cases in this category.
The columns \emph{StoryDroid, Fax, Random Delm, Delm} represent the number of passed cases for StoryDroid, Fax, Random Delm, and Delm, respectively.
According to the experimental results, Delm exhibits an improved performance by successfully addressing 74 out of the 81 cases in IntentBench, thereby achieving a success rate of 91.36\%. This outperforms the comparative systems, namely, Random Delm, which only successfully handled 70 cases, Fax, which dealt with 67 cases, and StoryDroid, which managed to pass 58 cases.
Further emphasizing Delm's efficacy, out of 12,070 identified intents within real-world apps, Delm successfully executed 3,003, a figure considerably greater than those attained by StoryDroid (1,935), Fax (2,072), and Random Delm (2,171).

Baseline tools, due to their deficiency in providing accurate local and global contextual support, do not measure up to the robust performance of Delm. 
When considering real-world cases, it's important to note the complex nature of the architecture and logic embedded within industrial apps. This complexity poses a more substantial challenge for context mock-ups compared to the cases in IntentBench, which lack business logic. As a result, a diminished success rate is observed for these real-world cases in contrast to the IntentBench cases.

In the results, we identify 7 instances in the IntentBench where Delm encounters a failure, all of which belong to the \emph{Object extra param} category. 
This category involves the transference of serialized Java objects to activities via intents, and this difficulty is not unique to Delm - all tools under consideration fail in this category.
In our large-scale context mock-up experiment, we also discover that in some industrial apps, the object parameter is used as a conduit between intents to relay dynamically acquired information and encrypted authentication data. To address such activities, these contexts must first be dynamically procured before we can mock up the other contexts.

While locating the source code of the passed object class is viable for tools such as Delm and RD, the extraction of the object properties encapsulated therein remains a formidable challenge. This can be attributed to the liberty developers possess in fully customizing object properties. Consequently, only simplistic 'toy' objects with simply fixed properties can be successfully extracted, thereby allowing a single case to pass. This scenario is also applicable to both StoryDroid and Fax.
Moreover, the current guidelines for the Activity Manager (AM) tool in ADB~\cite{adb} do not encompass instructions for passing serialized objects via intents, indicating that this limitation cannot be bypassed by any tool that relies on ADB.

Although the manual configuration of properties and information carried by these object parameters could serve as a possible solution, it is imperative to note that in order to mitigate potential security risks in apps, we advise against developers disclosing this information during testing.
Given these findings, the mock-up and investigation of object parameters in intent, particularly those concerning app security, and privacy, will feature prominently in the scope of our future work.

\subsubsection{Results of False Positives}


In our experiments, identifying a true positive crash hinges on satisfying two essential criteria simultaneously:
(1) Consistency of Stack Trace: Matching stack traces in both the modified and unmodified versions of the app indicate that the crash's root cause is inherent to the app's code, not a byproduct of experimental modifications. This consistency is crucial for asserting that the detected crash reflects a genuine flaw within the app.
(2) Reproducibility in the Unmodified App: A crash must be replicable under the same conditions, including the sequence of activities leading to the crash, in the unmodified version of the app. This requirement ensures that the crash originates from fundamental issues within the original codebase or its interactions with the operating system, libraries, or external services, rather than being induced by modifications or specific testing environments.

If two Android apps, one modified and the other unmodified, crash and produce the same stack trace, it only indicates that the crashes occurred at the same point in the code execution with the same call sequence leading up to the crash. However, this does not necessarily mean that the two apps have the same activity transition paths leading up to that point. Changing the activity transition path could still impact other aspects of the app's behavior, such as which activities are presented to the user or the state of the app when those activities are initiated. This divergence may induce crashes within the modified app that are not replicable in its original form.
Therefore, if a crash detected by Delm, shares the stack trace with a crash in the normal, unmodified app but fails to reproduce through standard navigation without deep links, it is classified as a false positive.

Table~\ref{tab:rq13} shows the results of the false positive experiment.
The columns \emph{\#Reported}, \emph{\#TP}, \emph{\#FP} and \emph{\#Uncover} represent numbers of total reported crashes, true positive crashes, false positive crashes and uncovered identified crashes for each tool.
Of the 52 crashes, Delm detects 28 crashes with 0 false positives.
22 crashes are detected by Fax, however, 4 of these are false positive crashes.
After reviewing the source code of apps, we think that the failure to mock up contexts is the cause of these false positive crashes.
Although the tool successfully launches the activity to be explored,  the launched activity will throw false positive exceptions like \emph{java.lang.NullPointerException} and \emph{IllegalArgumentException}, when it has to invoke these failure contexts.  

By combining the results in Table~\ref{tab:rq1} and Table~\ref{tab:rq13}, we can conclude that our tool can mock up more accurate activity context and decrease the likelihood of false positive results.

\begin{table}[]
\centering
\caption{Results of crash detection on Themis}
\small

\begin{tabular}{|c|c|c|c|c|}
\toprule
 Tool & \#Reported & \#TP & \#FP & \#Uncover \\
\midrule
Delm  & 28  & 28 & 0 & 24   \\
\rowcolor{gray!28}Fax       &   22  & 18     &  4     & 34  \\
\bottomrule
\end{tabular}
\label{tab:rq13}
\end{table}

\subsection{RQ2: Effectiveness of Activity Coverage}
\subsubsection{Dataset}
To further evaluate the capacity to launch activities in real-world apps, following related works~\cite{wang2018empirical, dong2020time, wang2021vet},
we randomly select 34 (50\%) out of 68 closed-source industrial apps from the IndustialApps~\cite{wang2018empirical} dataset.
Versions of all tested apps updated to February 2023.



\subsubsection{Experimental Setup}
In line with the time allocation adopted in related works~\cite{wang2018empirical, dong2020time}, each tool is allocated 3 hours for testing each app, along with 1 hour for preparation tasks such as model construction and instrumentation.
To ensure reliable results, we execute each app using each baseline tool a minimum of three times and select the highest activity coverage achieved~\cite{wang2018empirical}. The experiments are conducted on an Android Google Pixel 4 device with Android API level 28.

\subsubsection{Baselines}
To demonstrate the effectiveness of Delm's enhancement and compare its performance in terms of activity coverage, we have broadened our evaluation baselines. 
Alongside our established benchmarks StoryDroid, Fax, and Random Delm, we also introduce Monkey~\cite{monkey}, Fastbot~\cite{fastbot2}, Robo test~\cite{roboTest}, and Stoat~\cite{su2017guided}, serve as baselines against which Delm's performance can be evaluated.

Monkey~\cite{monkey} is a control baseline tool employed to showcase Delm's enhancement.
Fastbot\cite{cai2020fastbot, gu2019practical} is a model-based GUI testing tool.
It employs a multi-agent collaboration mechanism and reinforcement learning to speed up testing and jump out of the cyclic operation.
We select two typical testing tools Robo test~\cite{roboTest} and Stoat~\cite{su2017guided} as baselines.
Robo test is provided by Google Firebase test lab and widely used for close-sourced industry apps recently.
Stoat is a guided UI testing tool to perform stochastic model-based testing.


\begin{table}[]
\centering
\caption{The activity coverage of apps by different testing tools. Versions of all tested apps updated to February 2023.}
\begin{adjustbox}{max width=\linewidth}
\begin{tabular}{|l|l|l|cccccccc|}
    \toprule
 \multirow{2}{*}{ID} & \multirow{2}{*}{App Name} & \multirow{2}{*}{\#A} & \multicolumn{8}{c|}{Activity Coverage (\%)}  \\
    & &   & SD & Fax & St & RB & FB  & Mo & RD & Delm                         \\
    \midrule
    \rowcolor{gray!25}1&WhatsApp & 294 & 1 & 1 & 1 & 1 & 1 & 1 & 1 & \textbf{7}  \\
    2&EzFileExplorer & 228 & 3 & 4 & 1 & 1 & 12 & 4 & 11 & \textbf{14} \\
    \rowcolor{gray!25}3&Redfin & 145 & 6 & 10 & 8 & 10 & 15 & 15 & 14 & \textbf{24}  \\ 
    4&Wish  & 137 & 9 & 6 & 11 & 12 & 16 & 13 & 13 & \textbf{17} \\
    \rowcolor{gray!25}5&Tab  & 108 & 6 & 5 & 8 & 6 & 6 & 7 & 9 & \textbf{22}  \\ 
    6&NewsBreak & 107 & 10 & 10 & 11 & 8 & 16 & 17 & 16 & \textbf{27} \\  
    \rowcolor{gray!25}7&Spotify & 106 & 6 & 8 & 5 & 4 & 13 & 6 & 13 & \textbf{23} \\  
    8&Quizlet & 87 & 14 & 24 & 15 & 18 & 21 & 14 & 24 & \textbf{34} \\  
    \rowcolor{gray!25}9&Zillow & 85 & 22 & 22 & 16 & 13 & 44 & 24 & 45 & \textbf{49} \\  
    10&AmazonPrimeVideo & 78 & 1 & 4 & 1 & 1 & 10 & 4 & 4 & \textbf{19} \\  
    \rowcolor{gray!25}11&AudifyMusicPlayer  & 72 & 7 & 11 & 11 & 10 & 27 & 24 & 28 & \textbf{36}\\  
    12&LineCamera & 68 & 21 & 16 & 19 & 16 & 21 & 21 & 16 & \textbf{32} \\  
    \rowcolor{gray!25}13&Alltrails & 67 & 15 & 15 & 16 & 16 & 31 & 25 & 31 & 31 \\  
    14&Zello & 65 & 8 & 13 & 25 & 22 & 31 & 31 & 14 & \textbf{45} \\  
    \rowcolor{gray!25}15&YouTube & 64 & 2 & 9 & 5 & 5 & 13 & 13 & 8 & \textbf{28} \\  
    16&Shazam & 63 & 10 & 11 & 8 & 8 & 35 & 30 & 25 & \textbf{57} \\  
    \rowcolor{gray!25}17&Domain & 60 & 18 & 15 & 15 & 15 & \textbf{20} & 15 & 15 & 15 \\  
    18&Trivago & 58 & 52 & 55 & 47 & 36 & 62 & 52 & 64 & \textbf{69} \\  
    \rowcolor{gray!25}19&Duolingo & 57 & 25 & 28 & 19 & 19 & 30 & 25 & 37 & \textbf{39} \\  
    20&AustraliaPost & 56 & 20 & 25 & 36 & 32 & 30 & 30 & 25 & \textbf{43} \\  
    \rowcolor{gray!25}21&Smart Cleaner Pro & 55 & 7 & 7 & 5 & 7 & 16 & 7 & 11 & \textbf{18} \\  
    22&TuneinRadio & 53 & 11 & 9 & 8 & 6 & 11 & 11 & 17 & \textbf{26} \\  
    \rowcolor{gray!25}23&Realtor & 52 & 13 & 13 & 13 & 15 & 15 & 19 & 17 & \textbf{31} \\  
    24&\small LoseWeightAppForMen & 51 & 24 & 10 & 24 & 20 & 16 & 18 & 10 & \textbf{27} \\  
    \rowcolor{gray!25}25&BBC News & 51 & 12 & 16 & 22 & 25 & 24 & 16 & 18 & \textbf{30} \\  
    26&IbisPainX & 50 & 10 & 14 & 10 & 12 & 14 & 14 & 14 & 14\\  
    \rowcolor{gray!25}27&Otter & 46 & 13 & 22 & 13 & 9 & 26 & 17 & 20 & \textbf{33} \\  
    28&Etsy & 40 & 8 & 8 & 8 & 8 & 25 & 5 & 15 & \textbf{38} \\  
    \rowcolor{gray!25}29&Podcast Republic & 40 & 13 & 15 & 13 & 10 & 35 & 10 & 5 & \textbf{45} \\  
    30&Flipboard & 39 & 33 & 36 & 13 & 10 & 46 & 31 & 33 & \textbf{49} \\  
    \rowcolor{gray!25}31&Excel & 31 & 19 & 19 & 16 & 16 & 26 & 19 & 19 & \textbf{26}\\  
    32&FiltersForSelfie & 27 & 19 & 30 & 15 & 11 & 37 & 30 & 37 & \textbf{48} \\  
    \rowcolor{gray!25}33&Evernote & 24 & 25 & 20 & 17 & 13 & 33 & 17 & 33 & \textbf{42} \\  
    34&FloorPlanGenerator & 16 & 44 & 44 & 44 & 44 & 50 & 44 & 44 & \textbf{50} \\ 
    \midrule
    \rowcolor{gray!25}~& Average & 75.88 & 11.07 & 14.85 & 11.36 & 10.62 & 19.24 & 14.85 & 17.02 & \textbf{27.20} \\  

\bottomrule
\end{tabular}
\end{adjustbox}
\label{tab:actCov}
\end{table}

\subsubsection{Results}
Table~\ref{tab:actCov} shows the results of activity coverage in real-world apps.
Column \emph{\#A} represents the number of activities in this app.
\emph{SD, St, RB, FB, Mo} and \emph{RD} are short for StoryDroid, Stoat, Robo test, Fastbot Monkey and Random Delm.
The activity coverage of Delm in industrial apps reaches 27.20\%,  41.37\% higher than the best baseline Fastbot (19.24\%), significantly ahead of Random Delm's 17.02\%, Monkey's 14.85\%, Fax's 14.85\%, Stoat's 11.36\%, StoryDroid's 11.07\%, and Robo test's 10.62\%.
The results from Monkey and Delm show a noticeable enhancement by Delm for the exploration coverage in the complicated industrial apps, with a 83.16\% increase.
These results reveal that traditional tools have a significant deal of potential for improvement with the enhancement of Delm.
As indicated in Table ~\ref{tab:actCov}, the average activity coverage for Random Delm experienced a decline from 27.2\% with the original Delm to 17.2\%, as noted in the RD and Delm columns, corresponding to a reduction of 36.76\%. This data highlights the significant impact of our guided exploration algorithm, particularly its capacity for preserving global data and activity stacks and steering the testing tool toward activities yet to be explored. The observed decrease in activity coverage effectively illustrates the algorithm's critical role in optimizing GUI testing efficacy.

The Mann–Whitney U-test, an essential non-parametric statistical method, is employed to compare two independent samples for differences in their distribution~\cite{mcknight2010mann}. 
We applied the Mann–Whitney U-test to rigorously evaluate the activity coverage of Delm against various baseline tools as part of our analysis for RQ2.
The outcomes reveal that Delm consistently outperforms the baseline tools across all comparisons, with statistical significances supported by p-values of p < 0.001 for comparisons against StoryDroid, Fax, Stoat, RoboTest, Monkey, and Random Delm, and p = 0.006 against Fastbot.
These results, all of which are significantly below the conventional significance threshold of 0.05, unequivocally indicate that Delm's performance in activity coverage is statistically superior to that of the baseline tools, underscoring its effectiveness in enhancing activity coverage in app testing scenarios.

\begin{figure}
    \centering
    \includegraphics[width=0.8\linewidth]{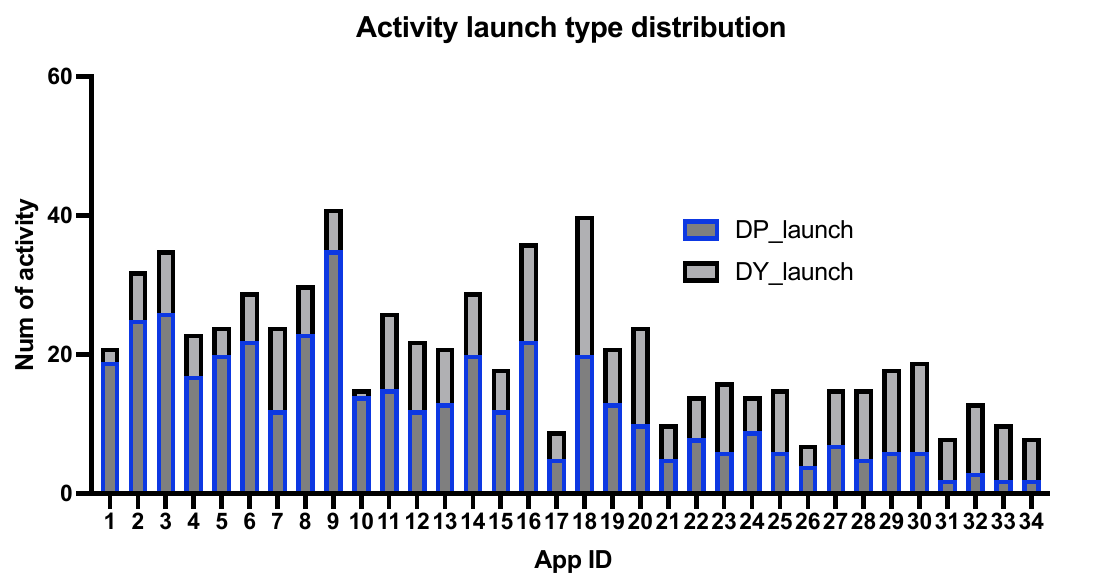}
    \caption{Activity launch type distribution. DY\_launch stands for dynamically launched activities, whereas DP\_launch refers to activities launched via deep links.
    }
    \label{fig:launchDis}
    \setlength{\belowcaptionskip}{5pt}
\end{figure}

To examine the impact of deep links on different apps, we analyze the activity launch type of covered activities in the 34 industrial apps. Figure~\ref{fig:launchDis} shows the distribution of activity launch types among the covered activities in the 34 apps, where the x-axis corresponds to the IDs of the apps listed in Table~\ref{tab:actCov}. We use the term \emph{DY\_launch} to represent dynamically launched activities and \emph{DP\_launch} to denote activities launched via deep links. As shown in Table~\ref{tab:actCov}, the total number of activities decreases from ID 1 to 34. We observe that the rate of deep link-based launch decreases steadily as the number of activities in the apps decreases, whereas the number of dynamically launched activities gradually increases. This suggests that more complex apps require greater enhancement from Delm when being tested.



Although the gaps for Fax, Random Delm, and Delm in Table~\ref{tab:rq1} are not significant, they are apparent in Table~\ref{tab:actCov}. Typically, more complicated apps should rely more on deep link-based launch and guided exploration. Our tool effectively alleviates this challenging issue by guiding exploration and mocking context in complex industrial apps.

\begin{figure}[htbp]
    \centering
    \includegraphics[width=0.7\linewidth]{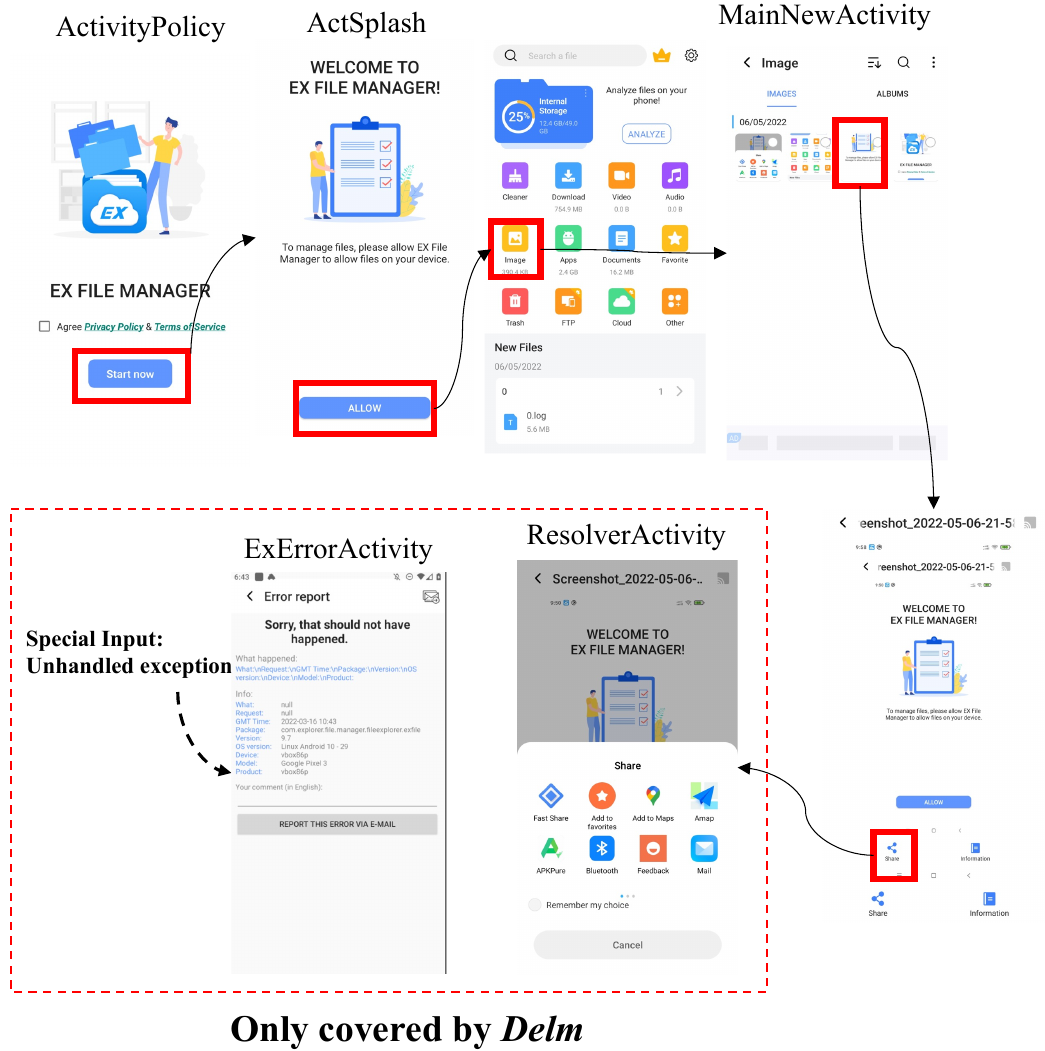}
    \caption{An example of activities covered by Delm and other tools in 'EZ File Explorer'.
    \emph{ExErrorActivity} and \emph{ResolverActivity} are both only covered by Delm.} 
    \label{fig:openCases}
\end{figure}

Figure~\ref{fig:openCases} presents illustrative instances within the real-world app \emph{EZ File Explorer}, facilitating a comparative analysis of activity coverage between Delm and baselines. 
The activities \emph{ResolverActivity} and \emph{ExErrorActivity} can only be covered by Delm.

To access \emph{ResolverActivity}, the baseline tools must transit via \emph{ActivityPolicy} -> \emph{ActSplash} -> \emph{MainNewActivity} and trigger the required events in the entire transition path. However, we observe that the entry points of the ResolverActivity are covered by numerous UI components, necessitating specific pre-operations for their direct presentation within the current interface. This architectural complexity often leads baseline tools to overlook the entry points of the activity, consequently bypassing all subsequent GUI pages associated with the \emph{ResolverActivity}. However, Delm discerns during the static analysis phase that \emph{ResolverActivity} can be launched directly without any additional ICC messages. This strategic insight allows Delm to bypass preliminary operations, directly triggering the intent associated with \emph{ResolverActivity}. This approach effectively alleviates the challenge of reaching activities that require navigating through extensive pre-subpaths or those whose entry points are not readily apparent.

Similar to the \emph{ResolverActivity}, after analyzing the decompiled code of the app, Delm finds that when the app crashes or fails due to uncontrollable external factors, the app catches this exception in code and passes a primary extra parameter --$es\ extra\_key$ with value $other$ to launch \emph{ExErrorActivity}. Therefore, visiting the activity \emph{ExErrorActivity} requires a special input and a primary extra parameter to the app, but no baselines can accurately send the special inputs to trigger it, resulting in getting stuck there. Conversely, Delm triggers existing intents with pre-discovered parameters in \emph{ExErrorActivity} to reach it, leading to the further exploration of related GUI pages and subsequent activities in its transition path.
Note that, in this scenario, Delm does not require the mocking of unhandled exceptions objects but merely the simulation of a primary extra parameter, specifically --$es\ extra\_key$, assigned the value $other$.

From the above examples, we can see that compared to baselines, Delm obtains better code coverage for three reasons.
\begin{itemize}
    \item (1) Direct Activity Launch through the Deep Link:
Delm utilizes the deep link to bypass complex pre-operations, allowing the direct launching of activities that are otherwise challenging to access, thereby preventing the omission of hard-to-find activity entrances by the testing tool.
    \item (2) Special Input Requirements for Activity Triggering:
It addresses the challenge of triggering activities that require special inputs (e.g., unhandled exceptions) by employing a combination of static analysis and dynamic exploration algorithms. This enables Delm to identify and initiate such activities directly, a task that general testing tools struggle with.
    \item (3) Improvement in Navigating Activity Transition Paths:
In the app's activity transition paths, failing to access a specific activity likely leads to missing subsequent activities in that path, adversely affecting code coverage. Delm's method of merging static activity context mock-ups with dynamic guided exploration algorithms significantly boosts its ability to access and navigate through activities that are deeply embedded within the app structure.
\end{itemize}

\subsubsection{Limitation}
In our analysis, we also examine the failure cases of Delm and identify several common causes of missed activities.
Some special development GUI patterns in apps may have difficulty obtaining detailed UI information, which prevents Delm from monitoring the exploration process.
For example, Delm's coverage of the app Domain is lower than baselines, as most activities in Domain are embedded into one \emph{WebView} based activity that cannot require detailed GUI components information via the current techniques.
Second, we also find that some app developers will deliberately confuse the type and method of passing parameters, resulting in lower accuracy of intent information analysis.
Faced with these limitations, we recommend using Delm to analyze Java source code instead of APKs.
Our subsequent research will focus on developing better extraction techniques and designing algorithms to reduce reliance on metadata.

\subsection{RQ3 Effectiveness of App Method Coverage}
\label{sec:exp2}

To investigate Delm's performance on code coverage, 
following Fax~\cite{yan2020multiple} and Fastbot~\cite{fastbot2}, we run Delm and baselines on real-world apps to compare the method coverage and detected crashes of each tool.

\subsubsection{Baselines}
\label{sec:baselines}

All baselines from RQ2 are reused, with the exception of StoryDroid.
StoryDroid features an activity launch module that allows it to partially access app activities, aligning with RQ2's goal of comparing activity coverage among different tools. 
This functionality justifies StoryDroid's inclusion in RQ2, as its ability to navigate GUI activities is relevant for assessing activity coverage, a key metric for this research question.
However, RQ3 focuses on evaluating tools that offer broader capabilities in GUI testing, including crash detection. 
StoryDroid's limited functionality in this more specialized domain rendered it unsuitable for inclusion in the method coverage analysis of RQ3. 
This decision was made to ensure the relevance and appropriateness of the tools assessed in relation to the specific aims of RQ3.
This decision highlights the distinction between the objectives of RQ2 and RQ3, with StoryDroid fitting the criteria for the former but not the latter.

\subsubsection{Dataset}
We select more datasets to enhance the generality of the experiment.
We attempt to use ELLA~\cite{ella}, which is also used in Stoat and Fastbot, to instrument closed-source apps in RQ2's dataset to obtain method coverage.
However, only 10 apps are successfully instrumented after much effort.
We follow Fax and Stoat to randomly select 
another 10 out of 68 open-source apps from AndroTest~\cite{choudhary2015automated}, which is already used as the benchmark by many other Android testing tools~\cite{yan2020multiple, baek2016automated, dong2020time, su2017guided, dong2020time}.
The selection of a new dataset for our study is driven by two key factors. First, the verification of true positive crashes in closed-source industrial apps demands extensive engineering efforts, including detailed stack trace comparisons, smali code analysis after decompilation, and manual crash reproduction. In contrast, open-source apps in the AndroidTest benchmark have a documented history of verified crashes, enhancing the efficiency of our tool's crash detection evaluation while maintaining accuracy. Second, preliminary analysis shows that crashes in industrial apps selected for RQ2 are rare and often unique. To ensure a comprehensive evaluation, we integrate the AndroidTest benchmark, which covers a wider range of crash types, into our study, enabling a thorough assessment of our tool's performance.
As shown in Table~\ref{tab:coverage}, the first 10 apps in the benchmark  (from \emph{Duoingo} to \emph{Filter For Selfie}) are closed-source industrial apps and the next 10 apps are open-source apps.

\subsubsection{Experiment setup}
The hardware and timing settings for this experiment match those of the RQ2 experiment.
In the experiment, the interval between each testing event in each tool is set at 0.5s.
We followed previous works~\cite{gu2019practical,wang2018empirical} to give maximum 1 hour to Stoat for model construction and 0.5 hour to Delm for instrumentation, both of which are counted into the test time.
We pre-register and log in to all apps if needed before the experiment.
JaCoCo~\cite{JaCoCo} and Ella~\cite{ella} are used to evaluate the code coverage for open-source apps and closed-source apps, respectively.
For crash detection, we follow Stoat~\cite{su2017guided} and Fax~\cite{dong2020time} to search the error stack of unique crashes in the output of Logcat~\cite{logcat} to detect crashes.
Every detected crash must be checked and reproduced manually to avoid false positives.

\begin{table}[!ht]
    \centering
    \caption{Results of method coverage and crash detection}
    \begin{adjustbox}{max width=\linewidth}
    \begin{tabular}{|l|l|llllll|llllll|}
    \toprule
        \multirow{2}{*}{App} & \multirow{2}{*}{M (\#)} & \multicolumn{6}{c|}{Method (\%)} & \multicolumn{6}{c|}{Crash (\#)} \\ 
        ~& ~& Delm & FB & St & RB & Fax & Mo & Delm & FB & St & RB & Fax & Mo \\ 
        \midrule

        \rowcolor{gray!25}Duolingo & 63,303 & \textbf{31} & 28 & 25 & 29 & 24 & 25 & 3 & 3 & 0 & 2 & 2 & 1 \\
        Trivago & 50,879 & \textbf{15} & 14 & 11 & 12 & 14 & 12 &  0 & 0 & 0 & 0 & 0 & 0 \\     
        \rowcolor{gray!25}Excel & 48,849 & 18 & 18 & 12 & 16 & 18 & 18 & 0 & 0 & 0 & 0 & 0 & 0 \\       
        Wish & 48,207 & 17 & 17 & 14 & 14 & 15 & 15 & 2 & 2 & 1 & 0 & 0 & 0\\
        \rowcolor{gray!25}IbisPainX & 47,721 & 18 & 18 & 10 & 17 & 18 & 18 & 1 & 0 & 1 & 1 & 1 & 1 \\         
        Evernote & 45,880 & \textbf{21} & 17 & 10 & 11 & 13 & 14 & 0 & 0 & 0 & 0 & 0 & 0 \\
        \rowcolor{gray!25}Spotify & 44,556 & 14 & 14 & 13 & 13 & 13 & 13 & 0 & 0 & 0 & 0 & 0 & 0 \\
        Wattpad & 44,069 & \textbf{16} & 15 & 13 & 13 & 14 & 14 & 0 & 0 & 0 & 0 & 0 & 0 \\     
        \rowcolor{gray!25}Flipboard & 41,563 & \textbf{22} & 19 & 16 & 17 & 20 & 20 & 3 & 2 & 1 & 1 & 3 & 2\\  
        FiltersForSelfie & 17,145 & 17 & 17 & 11 & 12 & 14 & 15 & 0 & 0 & 0 & 0 & 0 & 0 \\

        \midrule
        \rowcolor{gray!25}MyExpenses & 58,389 & \textbf{35} & 31 & 22 & 28 & 23  & 23 & 3 & 2 & 3 & 2 & 1 & 1\\ 
        anyMemo & 56,474 & \textbf{19} & 16 & 11 & 13 & 17 & 17 & 10 & 10 & 13 & 9 & 12 & 12\\ 
        \rowcolor{gray!25}Amaze & 41,475 & 14 & 12 & 12 & 13 & 14 & 14 & 19 & 18 & 8 & 10 & 13 & 13\\
        Commons & 31,124 & \textbf{15} & 13 & 13 & 12 & 13 & 12 & 5 & 0 & 1 & 0 & 1 & 1\\
        
         \rowcolor{gray!25}nextcloud & 10,402 & \textbf{22} & 19 & 18 & 13 & 16 & 17 & 5 & 1 & 0 & 3 & 2 & 1\\ 
        
        K9mail & 6733 & \textbf{56} & 53 & 32 & 47 & 45 & 45 & 14 & 14 & 13 & 10 & 10 & 10\\

         \rowcolor{gray!25}sannity & 3124 & \textbf{29} & 26 & 19 & 23 & 24 & 25 & 3 & 2 & 3 & 1 & 3 & 2\\
        hotdeath & 1203 & \textbf{72} & 70 & 66 & 63 & 67 & 66 & 1 & 1 & 2 & 1 & 1 & 0\\
         \rowcolor{gray!25}zooborns & 817 & \textbf{38} & 33 & 31 & 33 & 36 & 34 & 2 & 2 & 2 & 1 & 2 & 2\\ 
        aCal & 462 & \textbf{36} & 33 & 21 & 27 & 30 & 30 & 7 & 6 & 5 & 4 & 5 & 5\\
      
        \midrule
        \rowcolor{gray!25} Average & 33,118 & \textbf{21.13} & 18.97 & 14.69 & 16.91 & 17.38 & 17.39 & - & - & - & - & - & -\\ 
        Summation & ~ & - & - & - & - & - &- & \textbf{78} & 63 & 53 & 45 & 56 & 51 \\        
        
        \bottomrule

    \end{tabular}
    \end{adjustbox}
    \label{tab:coverage}
\end{table}

\subsubsection{Results}
Table~\ref{tab:coverage} shows the achieved method coverage and detected crashes by each tool in this benchmark.
\emph{\#M} means the number of methods in this app.
Table~\ref{tab:coverage} shows the method coverage results of five baselines and Delm.
It can be seen that Delm achieves the highest method coverage 21.13\%, compared 18.97\% for Fastbot, 17.39\% for Monkey, 17.38\% for Fax, 16.91\% for Robo and 14.69\% for Stoat. 
Delm is 11.39\% higher than the best baseline Fastbot, and given that the average number of methods in this dataset is 33,118, a 11.39\% lead is a significant improvement in code coverage.
The results from Delm and Monkey also prove the effectiveness of Delm's code coverage enhancement for existing testing tools.
Based on our observations, the reason behind achieving higher code coverage lies in our ability to explore a greater number of GUI states and activities, allowing us to thoroughly test the underlying logical code of these GUI states.
Our strengths in context mock-up and activity launch play a crucial role in enabling dynamic exploration, forming a solid foundation for attaining higher code coverage.

After the manual check, Delm also detected more unique crashes (78) on dynamic launched activities than Fastbot (63), Stoat (53), Robo (45), Fax (56) and Monkey (51).
Of all the reported crashes in our tool, we find no false positive crashes.
It achieves a 23.81\% improvement over the best baseline and an increase of 15 crashes.
The top three crash types among Delm's 78 identified crashes are \emph{NullPointerException}, \emph{IllegalArgumentException} and \emph{IndexOutofBoundsExeption}, accounting for 60, 9, and 3 crashes respectively.

\subsection{Ablation Study}

In the landscape of automated GUI testing, understanding the discrete contribution of individual components to the overall effectiveness is paramount. Our ablation study meticulously deconstructs our integrated approach, which comprises activity contexts mock-up, dynamic loop detection, and guided exploration algorithm, to discern the value each component adds to method coverage and crash detection capabilities.
This study is vital for both the validation of our approach and the identification of avenues for refinement in automated GUI testing methodologies.

\subsubsection{Dataset}

The study employs a dataset consisting of five open-source apps, selected randomly from the RQ3 dataset to represent a wide array of sizes, complexities, and app categories. This strategic selection underpins the generalizability of our study's findings, reinforcing the adaptability of our testing approach across diverse software ecosystems.

\subsubsection{Ablation Configurations}

Three ablation configurations were established to assess the impact of each individual component within our approach:
\begin{itemize}
    \item (1) Without Activity Contexts Mock-up (\emph{WACM}): This setup examines Delm's performance without the precise simulation of activity contexts to highlight the critical role of context accuracy. This configuration is referred to as \emph{WACM} in Table~\ref{tab:ablation}.
    \item (2) Without Dynamic Loop Detection (\emph{WDLD}): This configuration assesses Delm's functionality in the absence of dynamic loop detection mechanisms, underscoring the importance of avoiding redundant explorations for efficient GUI navigation. It is abbreviated as \emph{WDLD} in Table~\ref{tab:ablation}.
    \item (3) Without Guided Exploration Algorithm (\emph{WGEA}): This condition explores Delm's effectiveness without the guided exploration algorithm, focusing on its influence on GUI navigation and the overall testing coverage and crash detection. This scenario is labeled as \emph{WGEA} in Table~\ref{tab:ablation}.
\end{itemize}

The outcomes of method coverage and crash detection of Delm in RQ3 are retained to delineate the comparative effectiveness of the standard Delm configuration against the ablated versions.

\begin{table}[!ht]
    \centering
    \caption{Results of Ablation Study}
    \begin{adjustbox}{max width=\linewidth}
    \begin{tabular}{|l|l|llll|llll|}
    \toprule
        \multirow{2}{*}{App} & \multirow{2}{*}{M (\#)} & \multicolumn{4}{c|}{Method (\%)} & \multicolumn{4}{c|}{Crash (\#)} \\ 
        ~& ~& Delm & WACM & WDLD & WGEA & Delm & WACM & WDLD & WGEA \\ 
        \midrule
        anyMemo & 56,474 & \textbf{19} & 8 & 15 & 11 & 10 & 3 & 4 & 4 \\ 
        \rowcolor{gray!25}Amaze & 41,475 & 12 & 4 & 8  & 6  & 19 & 6 & 8 & 6\\
        
        K9mail & 6733 & \textbf{56} & 33 & 42 & 35  & 14 & 7 & 11  & 7\\

         \rowcolor{gray!25}sannity & 3124 & \textbf{29} & 16 & 20 & 19 & 3 & 2 & 2 & 2\\
        hotdeath & 1203 & \textbf{72} & 60 & 66 & 63 & 1 & 0 & 0 & 0\\
      
        \midrule
        \rowcolor{gray!25} Average & 21,801.8 & \textbf{37.6} & 24.2 & 30.2 & 26.8 & - & - & - & -\\ 
        Summation & ~ & - & - & - &- & \textbf{47} & 18 & 25 & 19 \\        
        
        \bottomrule

    \end{tabular}
    \end{adjustbox}
    \label{tab:ablation}
\end{table}

\subsubsection{Results}
Table~\ref{tab:ablation} presents the outcomes of our ablation study, elucidating the integral role of each component within our testing methodology.
The experimental results confirm the significant impact of intent context mock-up on method coverage. 
Specifically, as indicated in the WACM columns of Table~\ref{tab:ablation}, method coverage notably declined from 37.6\% to 24.2\% in the absence of this feature. Modern apps frequently hide certain activities or interfaces behind intricate user interface elements, which pose a challenge for activation via mere random exploration. The mock-up technology serves as a pivotal mechanism for uncovering these concealed functionalities, thereby augmenting the tool's capacity for achieving exhaustive coverage. The absence of this capability, as evidenced in our experimental configuration without activity contexts mock-up, leads to suboptimal code coverage.

Furthermore, the WDLD columns in Table~\ref{tab:ablation} highlight a reduction in average coverage to 30.2\% when the dynamic loop detection mechanism is disabled. In such scenarios, the Delm tool exhibited a propensity for entering redundant cycles or endlessly iterating over a confined assortment of GUI pages, thereby inefficiently allocating testing resources and, as a consequence, restricting the breadth of test code coverage. This phenomenon accentuates the critical importance of loop detection in optimizing the efficiency and comprehensiveness of the testing endeavor.

Moreover, the absence of the guided exploration strategy, as reflected in the coverage reduction to 26.8\%, underscores the challenges faced in accessing GUI pages with complex triggering mechanisms. Despite the loop detection mechanism's contribution to averting repetitive exploration of common interfaces, the tool struggled to penetrate deeper into the application to interact with sophisticated GUI elements. This limitation is markedly pronounced in the exploration of advanced GUI pages or activities, resulting in a discernible decrement in code coverage and underscoring the indispensability of guided exploration in enabling thorough testing coverage.

Overall, our ablation study reveals a notable decline in method coverage across the ablation configurations—WACM, WDLD, and WGEA—by 35.6\%, 19.7\%, and 28.7\%, respectively. 
This effect is particularly pronounced in apps with extensive method counts, such as Anymemo and Amaze. 
Correspondingly, the detection of crashes in these configurations diminished significantly, from 47 to 18, 25, and 19 crashes, respectively. 
These findings elucidate that apps of higher complexity disproportionately rely on the integrated algorithms within our framework, with a particular emphasis on activity mock-ups and guided exploration algorithms. The examination of test cases pertinent to these configurations has yielded valuable insights into the distinct and crucial contributions of each algorithm, underscoring their indispensability for achieving comprehensive testing coverage and effective crash detection in complex app environments.

\section{Threats to Validity}

Several threats to the validity of our study and our approach, Delm, need to be taken into account.

\textbf{Repackaging limitations}: A primary threat to Delm's effectiveness stems from the limitations of the app packaging format. Given Delm's focus on Java source code, it may struggle with Android APK file processing. This constraint might cause issues if an app cannot be repackaged appropriately into APK files. Extending Delm's compatibility with various packaging formats could be a worthwhile focus for future research.

\textbf{Incomplete ATG}: The second threat is related to the completeness of the ATG. Since the ATG may not be entirely exhaustive, Delm could potentially miss some activity context. This incompleteness might lead to false positives concerning app crashes. Addressing this threat would require the development of more robust methods for generating comprehensive ATGs.



\textbf{False Positive Crash}: 
A significant internal threat to Delm's validity is potential false positive crashes. Mitigation measures include recommending developers analyze their own source code instead of released APKs and customizing app-specific trigger rules for GUI testing. Additionally, Delm employs strict trigger rules, rejecting external triggers for intents lacking accurate contexts to further avoid false positives.

\textbf{Limitation of the Mock-up}:
The methodology deployed in this study is limited by its reliance on static mock-ups for simulating app behaviors, a significant constraint when endeavoring to replicate the dynamic data behaviors integral to authentic app scenarios. 
This approach is strategically chosen to mitigate the risk of false positive crashes, a non-trivial concern in automated GUI testing. 
Acknowledgment of this limitation is essential for a nuanced interpretation of our results and a comprehensive assessment of the method's scalability. 
Additionally, the rule-based static activity context analysis method employed in our study may fail to detect or address some activity contexts that possess ambiguous definitions or value assignments. 
To mitigate the risk of subsequent false positive crashes, our current approach deliberately avoids mocking up these potentially hazardous activities. 
This cautious strategy, however, introduces a degree of randomness into our ability to consistently launch activities. 
We are aware of these limitations and the potential impact they may have on the validity of our results. 
In future work, we aim to surmount these obstacles by developing more sophisticated methods that can accurately emulate dynamic data, thereby enhancing the reliability and applicability of our research findings.

\textbf{Experimental Design Limitations}:
Our study's experimental design encompasses inherent limitations, particularly regarding the execution frequency of randomized algorithms. Arcuri and Briand highlight the importance of running each randomized algorithm across various artifacts multiple times (ideally n = 1000, or at a minimum, n = 10) to ensure statistical robustness and account for randomness~\cite{arcuri2014hitchhiker}. 
Our operational constraints, primarily the extensive duration required to execute a 34-app dataset through seven tools for three hours each, plus additional time allocated for statistical analysis and debugging, made such comprehensive execution unfeasible. 
Consequently, we chose to align our methodology with partial baselines by selecting the best outcome from three repetitions. This approach, while has been adopted in prior studies, may not fully adhere to the statistical rigor advised by Arcuri and Briand, potentially leading to optimistic assessments of the algorithm's effectiveness. 
Recognizing this, we highlight the importance of future research endeavors to engage in more detailed statistical analyses to better navigate the challenges posed by randomness in experimental setups.

\textbf{Sample Representativeness}:
The composition of our dataset, which includes 50\% closed-source apps, presents another limitation. To counteract this, we aime for a diverse and comprehensive selection of popular app types, endeavoring to minimize bias and ensure the study's relevance across varying software trends. The strategy to encompass a broad array of apps is devised to mitigate selection bias and bolster the representativeness of our research. Nonetheless, the selection methodology's limitations and its potential impact on the study's external validity are critical considerations, highlighting the need for cautious interpretation of our findings and their applicability to broader contexts.


\textbf{Object Parameters in ICC}:
Automated analysis of object parameters in Android's ICC communication faces challenges, especially with custom classes. These classes' complexity and the use of dynamic code complicate static analysis. 
The deep link reliance on URIs, suited for simple data types, limits complex object transmission. 
We address this by serializing object data into strings for URI embedding, albeit with inherent limitations like URI length and encoding challenges.
The variability in custom Java classes poses significant challenges, increasing the likelihood of false positives and necessitating manual verification. 

Despite these issues, object parameter transmission between intents is uncommon due to the extra development effort involved. In light of these considerations, Delm adopts a pragmatic approach, emphasizing deterministic constant parameters while relegating the analysis of object parameters to manual inspection. This strategy is employed selectively, particularly in instances where the constant value within an object parameter is explicitly identifiable. This methodological choice represents a calculated compromise aimed at optimizing the balance between the scope of automated analysis and the precision of the resultant findings, thereby ensuring the reliability of Delm's performance.
Future research will delve into exploring more efficient methods for passing object parameters, aiming to enhance the robustness and accuracy of automated analysis in the context of Android ICC.

\section{Related Work}
This section presents a discussion on the related work pertaining to three main areas: deep links, activity context analysis, and Android GUI exploration and testing. 

\subsection{Deep Link}
Several studies have explored aiding developers in integrating deep links into apps. Notably, Azim et al.\cite{azim2016ulink} developed uLink, a library facilitating app page encoding into links, and Aladdin\cite{ma2018aladdin} automates deep link generation in app codebases. Furthermore, Tang et al.~\cite{tang2020all} examined deep link hijacking, identifying numerous vulnerabilities in contemporary apps. However, no approach has yet focused on deep links for GUI exploration in app testing. Our work aims to fill this niche, leveraging deep links to enhance app UI testing.

\subsection{Activity Context Analysis}
Static analysis is a common approach for activity context extraction in Android apps, as seen in tools like Fax~\cite{yan2020multiple} and methods from Lai et al.~\cite{lai2019goal}. Both generate test path scripts and state transition graphs to assist in triggering app functionalities. IC3 \cite{octeau2015composite, tian2018poster} introduces COAL language for modelling inter-component communication in Android activities. Despite these advancements, current static analysis methods still struggle with accurately extracting intent information for deep links to activities. Our study aims to overcome these limitations, enhancing static analysis efficacy in extracting deep link intent information.

\subsection{GUI Exploration and Testing}

Testing tools come in various flavors, including model-based, probability-based, symbolic execution-based, and machine learning-based methods. 

Model-based GUI testing tools analyze source code or currently displayed dynamic GUIs to build a model that will guide app testing~\cite{mao2016sapienz, su2017guided, andrieu2003introduction, hu2024iOS, hu2019code, gu2019practical, yan2020multiple, chen2021my, liu2022dalt, auer2023android}.
Sappienz~\cite{mao2016sapienz} pre-defines patterns of lower-level events to capture complicated interactions and test event paths for higher code coverage in less time cost.
Inspired by Sappienz, Stoat~\cite{su2017guided} builds a stochastic model and UI state transition model by static analysis and dynamic exploration.
Then, it adapts Gibbs sampling~\cite{andrieu2003introduction} with a stochastic model to guide the test path.
However, as apps become more complex, it is difficult to extract enough exploration paths from static analysis.
Recently, Ape~\cite{gu2019practical} represents the dynamic model as a decision tree and continuously tunes using the feedback obtained during testing.
Fax, designed for Android, generates multi-entry tests by analyzing source code to construct activity contexts and orchestrate tests based on activity importance~\cite{yan2020multiple}. However, it lacks dynamic testing enhancement algorithms, failing to follow the app's native activity flow or consider inter-activity contexts, thus raising false positive risks.
Aimdroid utilizes a multi-level testing framework optimizing activity transitions via models and reinforcement learning, focusing on singular activity exploration to avoid excess transitions~\cite{gu2017aimdroid}. Contrastingly, Delm bypasses conventional transitions, directly enhancing activity, method coverage, and crash detection by deep links and activity context mock-up. This highlights both tools' contributions to automated Android testing through different strategies.
DALT focuses on using static analysis to extract intent constraints for better GUI testing and bug detection, requiring open-source apps for access to intent details~\cite{liu2022dalt}. In contrast, Delm also gathers information from the runtime, offering a more practical approach for testing and exploration, including deep link support for direct GUI navigation, thus improving coverage and crash detection.
Auer et al.~\cite{auer2023android} also enhances Android app testing by balancing user input with fuzzy intents for broader testing efficiency. This balanced method boosts testing by leveraging the complementary strengths of user-driven events and intent-based navigation, but like DALT, it also requires the app to be open-sourced.
Delm utilizes deep links and designs corresponding dynamic algorithms to direct tests toward specific, often overlooked activities, relatively more concise.
Although the model-based approach adapts different models for testing, the core idea is to build a model for guiding the app testing to increase the code coverage.
However, powered by the Android anti-decompilation technology, it is hard to build an accurate testing model to guide most testing paths, especially for deep stack activities.
Most model-based testing tools stubbornly loop among common app GUIs, which is extremely inefficient.

Probability-based tools randomly generate events on the Android GUI, like clicks, slides, etc., and then apply them to random UI components~\cite{hu2011automating, mahmood2012whitebox, monkey}.
AppDoctor~\cite{hu2014efficiently} and Monkey~\cite{monkey} randomly invoke event handlers on the app screen to test the robustness of Android apps.
Furthermore, Dynodroid~\cite{machiry2013dynodroid} applies pre-defined heuristics for dynamic GUI testing to enhance random testing. 
Overall, random-based methods are not stable. 
To test activities as many as possible,  users have to set aside a large test time to ensure the adequacy of the test.

Symbolic execution-based testing approaches utilize symbolic execution techniques to exhaustively explore possible program execution paths in testing, which can simultaneously explore multiple program paths with various execution contexts~\cite{anand2012automated, jensen2013automated, mirzaei2012testing, gao2018android}.
However, most of them are very heavy, without the capability to scale up to complicated Android apps.

There are some machine learning-based methods such as recurrent neural network and reinforcement learning, which mimic human behaviors to explore apps, resulting in relatively higher testing coverage with less time cost ~\cite{li2019humanoid,pan2020reinforcement,liu2021promal, hu2024pairwise}.
Several studies have focused on enhancing the efficiency of exploration through optimization of specific UI components, such as text inputs \cite{liu2024testing, liu2023fill}, particular processes like GUI rendering \cite{feng2023efficiency}, and distinct platforms \cite{hu2023pairwise, hu2024TVGUI}. However, while these efforts contribute valuable insights for targeted scenarios, they exhibit a limited scope by primarily addressing specific types, functionalities, or platforms of apps. This focus inherently restricts their generalizability across the broader spectrum of application development.
When dealing with complex industrial Android apps requiring longer testing pathways, existing tools cannot reliably locate deep-stack activities or activities requiring complicated inputs, resulting in limited coverage.
In contrast, our tool uses deep links to launch activities when no dynamic paths are found to access these activities directly.

\section{Conclusion}
In this paper, we propose our deep link-enhanced GUI exploration approach for Android apps and develop the tool deep link-enhanced Monkey (Delm) to evaluate the effectiveness of our approach.
The tool consists of two stages: static analysis and guided exploration enhancement.
The evaluations demonstrate the effectiveness of Delm in improving activity context mock-up, activity coverage, method coverage, and crash detection.

Future work will attempt to utilize computer vision techniques to extract information from the GUI state and lessen reliance on ADB tools. 
Meanwhile, we will attempt to guide the app to leave the exploration loop on its own, without the use of external trigger intents.

\bibliographystyle{ACM-Reference-Format}
\bibliography{reference}


\begin{thebibliography}{83}


\ifx \showCODEN    \undefined \def \showCODEN     #1{\unskip}     \fi
\ifx \showDOI      \undefined \def \showDOI       #1{#1}\fi
\ifx \showISBNx    \undefined \def \showISBNx     #1{\unskip}     \fi
\ifx \showISBNxiii \undefined \def \showISBNxiii  #1{\unskip}     \fi
\ifx \showISSN     \undefined \def \showISSN      #1{\unskip}     \fi
\ifx \showLCCN     \undefined \def \showLCCN      #1{\unskip}     \fi
\ifx \shownote     \undefined \def \shownote      #1{#1}          \fi
\ifx \showarticletitle \undefined \def \showarticletitle #1{#1}   \fi
\ifx \showURL      \undefined \def \showURL       {\relax}        \fi
\providecommand\bibfield[2]{#2}
\providecommand\bibinfo[2]{#2}
\providecommand\natexlab[1]{#1}
\providecommand\showeprint[2][]{arXiv:#2}

\bibitem[ell(2016)]%
        {ella}
 \bibinfo{year}{2016}\natexlab{}.
\newblock \bibinfo{title}{ELLA}.
\newblock \bibinfo{howpublished}{\url{https://github.com/saswatanand/ella}}.
\newblock
\newblock
\shownote{Accessed: 2023-08-20}.


\bibitem[all(2023)]%
        {alltrails}
 \bibinfo{year}{2023}\natexlab{}.
\newblock \bibinfo{title}{Alltrails}.
\newblock
  \bibinfo{howpublished}{\url{https://play.google.com/store/apps/details?id=com.alltrails.alltrails&hl=en_US&gl=US}}.
\newblock
\newblock
\shownote{Accessed: 2023-08-20}.


\bibitem[And(2023a)]%
        {Android10}
 \bibinfo{year}{2023}\natexlab{a}.
\newblock \bibinfo{title}{Android 10}.
\newblock
  \bibinfo{howpublished}{\url{https://developer.android.com/about/versions/10/highlights}}.
\newblock
\newblock
\shownote{Accessed: 2023-05-02}.


\bibitem[adb(2023)]%
        {adb}
 \bibinfo{year}{2023}\natexlab{}.
\newblock \bibinfo{title}{Android Adb}.
\newblock
  \bibinfo{howpublished}{\url{https://developer.android.com/studio/command-line/adb}}.
\newblock
\newblock
\shownote{Accessed: 2023-07-04}.


\bibitem[bun(2023)]%
        {bundle}
 \bibinfo{year}{2023}\natexlab{}.
\newblock \bibinfo{title}{Android Bundle}.
\newblock
  \bibinfo{howpublished}{\url{https://developer.android.com/reference/android/os/Bundle?hl=en}}.
\newblock
\newblock
\shownote{Accessed: 2023-08-09}.


\bibitem[And(2023b)]%
        {AndroidContext}
 \bibinfo{year}{2023}\natexlab{b}.
\newblock \bibinfo{title}{Android Context}.
\newblock
  \bibinfo{howpublished}{\url{https://developer.android.com/reference/android/content/Context}}.
\newblock
\newblock
\shownote{Accessed: 2023-07-30}.


\bibitem[int(2023a)]%
        {intent}
 \bibinfo{year}{2023}\natexlab{a}.
\newblock \bibinfo{title}{Android Intent}.
\newblock
  \bibinfo{howpublished}{\url{https://developer.android.com/guide/components/intents-filters}}.
\newblock
\newblock
\shownote{Accessed: 2023-08-09}.


\bibitem[moc(2023)]%
        {mockContext}
 \bibinfo{year}{2023}\natexlab{}.
\newblock \bibinfo{title}{Android Mock Context}.
\newblock
  \bibinfo{howpublished}{\url{https://developer.android.com/reference/android/test/mock/MockContext}}.
\newblock
\newblock
\shownote{Accessed: 2023-08-20}.


\bibitem[and(2023)]%
        {androidTesting}
 \bibinfo{year}{2023}\natexlab{}.
\newblock \bibinfo{title}{Android Testing}.
\newblock
  \bibinfo{howpublished}{\url{https://developer.android.com/training/testing/fundamentals/what-to-test}}.
\newblock
\newblock
\shownote{Accessed: 2023-08-09}.


\bibitem[dee(2023)]%
        {deeplink}
 \bibinfo{year}{2023}\natexlab{}.
\newblock \bibinfo{title}{Deep Link}.
\newblock
  \bibinfo{howpublished}{\url{https://developer.android.com/training/app-links/deep-linking}}.
\newblock
\newblock
\shownote{Accessed: 2023-08-09}.


\bibitem[ezf(2023)]%
        {ezfile}
 \bibinfo{year}{2023}\natexlab{}.
\newblock \bibinfo{title}{Ezfile}.
\newblock
  \bibinfo{howpublished}{\url{https://en.wikipedia.org/wiki/ES\_File\_Explorer}}.
\newblock
\newblock
\shownote{Accessed: 2023-08-09}.


\bibitem[goo(2023)]%
        {googleMapURL}
 \bibinfo{year}{2023}\natexlab{}.
\newblock \bibinfo{title}{GoogleMap Documents}.
\newblock
  \bibinfo{howpublished}{\url{https://developers.google.com/maps/documentation/urls/get-started}}.
\newblock
\newblock
\shownote{Accessed: 2023-06-21}.


\bibitem[ic3(2023)]%
        {ic3}
 \bibinfo{year}{2023}\natexlab{}.
\newblock \bibinfo{title}{IC3}.
\newblock \bibinfo{howpublished}{\url{https://github.com/siis/ic3}}.
\newblock
\newblock
\shownote{Accessed: 2023-08-20}.


\bibitem[int(2023b)]%
        {intentBench}
 \bibinfo{year}{2023}\natexlab{b}.
\newblock \bibinfo{title}{IntentBench Homepage}.
\newblock
  \bibinfo{howpublished}{\url{https://github.com/hanada31/Fax/tree/master/IntentBench}}.
\newblock
\newblock
\shownote{Accessed: 2023-08-20}.


\bibitem[JaC(2023)]%
        {JaCoCo}
 \bibinfo{year}{2023}\natexlab{}.
\newblock \bibinfo{title}{JaCoCo}.
\newblock \bibinfo{howpublished}{\url{https://www.eclemma.org/jacoco/}}.
\newblock
\newblock
\shownote{Accessed: 2023-08-20}.


\bibitem[Jav(2023)]%
        {JavaReflection}
 \bibinfo{year}{2023}\natexlab{}.
\newblock \bibinfo{title}{Java Reflection}.
\newblock
  \bibinfo{howpublished}{\url{https://www.oracle.com/technical-resources/articles/java/javareflection.html}}.
\newblock
\newblock
\shownote{Accessed: 2023-08-09}.


\bibitem[log(2023)]%
        {logcat}
 \bibinfo{year}{2023}\natexlab{}.
\newblock \bibinfo{title}{Logcat}.
\newblock
  \bibinfo{howpublished}{\url{https://developer.android.com/studio/command-line/logcat}}.
\newblock
\newblock
\shownote{Accessed: 2023-08-20}.


\bibitem[los(2023)]%
        {loseweight}
 \bibinfo{year}{2023}\natexlab{}.
\newblock \bibinfo{title}{Men Lose Weight}.
\newblock
  \bibinfo{howpublished}{\url{https://play.google.com/store/apps/details?id=menloseweight.loseweightappformen.weightlossformen&hl=en_US&gl=US}}.
\newblock
\newblock
\shownote{Accessed: 2023-08-20}.


\bibitem[mon(2023)]%
        {monkey}
 \bibinfo{year}{2023}\natexlab{}.
\newblock \bibinfo{title}{Monkey}.
\newblock
  \bibinfo{howpublished}{\url{https://developer.android.com/studio/test/monkeyrunner}}.
\newblock
\newblock
\shownote{Accessed: 2023-08-20}.


\bibitem[rob(2023)]%
        {roboTest}
 \bibinfo{year}{2023}\natexlab{}.
\newblock \bibinfo{title}{Robo Test}.
\newblock
  \bibinfo{howpublished}{\url{https://firebase.google.com/docs/test-lab/android/robo-ux-test}}.
\newblock
\newblock
\shownote{Accessed: 2023-08-20}.


\bibitem[Soo(2023)]%
        {Soot}
 \bibinfo{year}{2023}\natexlab{}.
\newblock \bibinfo{title}{Soot}.
\newblock \bibinfo{howpublished}{\url{https://github.com/soot-oss/soot}}.
\newblock
\newblock
\shownote{Accessed: 2023-08-09}.


\bibitem[act(2024)]%
        {actvityStack}
 \bibinfo{year}{2024}\natexlab{}.
\newblock \bibinfo{title}{Actvity Stack}.
\newblock
  \bibinfo{howpublished}{\url{https://developer.android.com/guide/components/activities/tasks-and-back-stack}}.
\newblock
\newblock
\shownote{Accessed: 2024-04-20}.


\bibitem[Al{\'e}groth et~al\mbox{.}(2015)]%
        {alegroth2015visual}
\bibfield{author}{\bibinfo{person}{Emil Al{\'e}groth}, \bibinfo{person}{Robert
  Feldt}, {and} \bibinfo{person}{Lisa Ryrholm}.}
  \bibinfo{year}{2015}\natexlab{}.
\newblock \showarticletitle{Visual gui testing in practice: challenges,
  problemsand limitations}.
\newblock \bibinfo{journal}{\emph{Empirical Software Engineering}}
  \bibinfo{volume}{20} (\bibinfo{year}{2015}), \bibinfo{pages}{694--744}.
\newblock


\bibitem[Anand et~al\mbox{.}(2012)]%
        {anand2012automated}
\bibfield{author}{\bibinfo{person}{Saswat Anand}, \bibinfo{person}{Mayur Naik},
  \bibinfo{person}{Mary~Jean Harrold}, {and} \bibinfo{person}{Hongseok Yang}.}
  \bibinfo{year}{2012}\natexlab{}.
\newblock \showarticletitle{Automated concolic testing of smartphone apps}. In
  \bibinfo{booktitle}{\emph{Proceedings of the ACM SIGSOFT 20th International
  Symposium on the Foundations of Software Engineering}}.
  \bibinfo{pages}{1--11}.
\newblock


\bibitem[Andrieu et~al\mbox{.}(2003)]%
        {andrieu2003introduction}
\bibfield{author}{\bibinfo{person}{Christophe Andrieu}, \bibinfo{person}{Nando
  De~Freitas}, \bibinfo{person}{Arnaud Doucet}, {and}
  \bibinfo{person}{Michael~I Jordan}.} \bibinfo{year}{2003}\natexlab{}.
\newblock \showarticletitle{An introduction to MCMC for machine learning}.
\newblock \bibinfo{journal}{\emph{Machine learning}} \bibinfo{volume}{50},
  \bibinfo{number}{1} (\bibinfo{year}{2003}), \bibinfo{pages}{5--43}.
\newblock


\bibitem[Arcuri and Briand(2014)]%
        {arcuri2014hitchhiker}
\bibfield{author}{\bibinfo{person}{Andrea Arcuri} {and} \bibinfo{person}{Lionel
  Briand}.} \bibinfo{year}{2014}\natexlab{}.
\newblock \showarticletitle{A hitchhiker's guide to statistical tests for
  assessing randomized algorithms in software engineering}.
\newblock \bibinfo{journal}{\emph{Software Testing, Verification and
  Reliability}} \bibinfo{volume}{24}, \bibinfo{number}{3}
  (\bibinfo{year}{2014}), \bibinfo{pages}{219--250}.
\newblock


\bibitem[Auer et~al\mbox{.}(2023)]%
        {auer2023android}
\bibfield{author}{\bibinfo{person}{Michael Auer}, \bibinfo{person}{Andreas
  Stahlbauer}, {and} \bibinfo{person}{Gordon Fraser}.}
  \bibinfo{year}{2023}\natexlab{}.
\newblock \showarticletitle{Android Fuzzing: Balancing User-Inputs and
  Intents}. In \bibinfo{booktitle}{\emph{2023 IEEE Conference on Software
  Testing, Verification and Validation (ICST)}}. IEEE, \bibinfo{pages}{37--48}.
\newblock


\bibitem[Azim et~al\mbox{.}(2016)]%
        {azim2016ulink}
\bibfield{author}{\bibinfo{person}{Tanzirul Azim}, \bibinfo{person}{Oriana
  Riva}, {and} \bibinfo{person}{Suman Nath}.} \bibinfo{year}{2016}\natexlab{}.
\newblock \showarticletitle{uLink: Enabling user-defined deep linking to app
  content}. In \bibinfo{booktitle}{\emph{Proceedings of the 14th Annual
  International Conference on Mobile Systems, Applications, and Services}}.
  \bibinfo{pages}{305--318}.
\newblock


\bibitem[Baek and Bae(2016)]%
        {baek2016automated}
\bibfield{author}{\bibinfo{person}{Young-Min Baek} {and}
  \bibinfo{person}{Doo-Hwan Bae}.} \bibinfo{year}{2016}\natexlab{}.
\newblock \showarticletitle{Automated model-based android gui testing using
  multi-level gui comparison criteria}. In
  \bibinfo{booktitle}{\emph{Proceedings of the 31st IEEE/ACM International
  Conference on Automated Software Engineering}}. \bibinfo{pages}{238--249}.
\newblock


\bibitem[Cai et~al\mbox{.}(2020)]%
        {cai2020fastbot}
\bibfield{author}{\bibinfo{person}{Tianqin Cai}, \bibinfo{person}{Zhao Zhang},
  {and} \bibinfo{person}{Ping Yang}.} \bibinfo{year}{2020}\natexlab{}.
\newblock \showarticletitle{Fastbot: A Multi-Agent Model-Based Test Generation
  System Beijing Bytedance Network Technology Co., Ltd.}. In
  \bibinfo{booktitle}{\emph{Proceedings of the IEEE/ACM 1st International
  Conference on Automation of Software Test}}. \bibinfo{pages}{93--96}.
\newblock


\bibitem[Chen et~al\mbox{.}(2021b)]%
        {chen2021my}
\bibfield{author}{\bibinfo{person}{Qiuyuan Chen}, \bibinfo{person}{Xin Xia},
  \bibinfo{person}{Han Hu}, \bibinfo{person}{David Lo}, {and}
  \bibinfo{person}{Shanping Li}.} \bibinfo{year}{2021}\natexlab{b}.
\newblock \showarticletitle{Why my code summarization model does not work: Code
  comment improvement with category prediction}.
\newblock \bibinfo{journal}{\emph{ACM Transactions on Software Engineering and
  Methodology (TOSEM)}} \bibinfo{volume}{30}, \bibinfo{number}{2}
  (\bibinfo{year}{2021}), \bibinfo{pages}{1--29}.
\newblock


\bibitem[Chen et~al\mbox{.}(2021a)]%
        {chen2021accessible}
\bibfield{author}{\bibinfo{person}{Sen Chen}, \bibinfo{person}{Chunyang Chen},
  \bibinfo{person}{Lingling Fan}, \bibinfo{person}{Mingming Fan},
  \bibinfo{person}{Xian Zhan}, {and} \bibinfo{person}{Yang Liu}.}
  \bibinfo{year}{2021}\natexlab{a}.
\newblock \showarticletitle{Accessible or Not An Empirical Investigation of
  Android App Accessibility}.
\newblock \bibinfo{journal}{\emph{IEEE Transactions on Software Engineering}}
  (\bibinfo{year}{2021}).
\newblock


\bibitem[Chen et~al\mbox{.}(2019)]%
        {chen2019storydroid}
\bibfield{author}{\bibinfo{person}{Sen Chen}, \bibinfo{person}{Lingling Fan},
  \bibinfo{person}{Chunyang Chen}, \bibinfo{person}{Ting Su},
  \bibinfo{person}{Wenhe Li}, \bibinfo{person}{Yang Liu}, {and}
  \bibinfo{person}{Lihua Xu}.} \bibinfo{year}{2019}\natexlab{}.
\newblock \showarticletitle{Storydroid: Automated generation of storyboard for
  Android apps}. In \bibinfo{booktitle}{\emph{2019 IEEE/ACM 41st International
  Conference on Software Engineering (ICSE)}}. IEEE, \bibinfo{pages}{596--607}.
\newblock


\bibitem[Choudhary et~al\mbox{.}(2015)]%
        {choudhary2015automated}
\bibfield{author}{\bibinfo{person}{Shauvik~Roy Choudhary},
  \bibinfo{person}{Alessandra Gorla}, {and} \bibinfo{person}{Alessandro Orso}.}
  \bibinfo{year}{2015}\natexlab{}.
\newblock \showarticletitle{Automated test input generation for android: Are we
  there yet?(e)}. In \bibinfo{booktitle}{\emph{2015 30th IEEE/ACM International
  Conference on Automated Software Engineering (ASE)}}. IEEE,
  \bibinfo{pages}{429--440}.
\newblock


\bibitem[Dashevskyi et~al\mbox{.}(2018)]%
        {codeCoverage}
\bibfield{author}{\bibinfo{person}{Stanislav Dashevskyi}, \bibinfo{person}{Olga
  Gadyatskaya}, \bibinfo{person}{Aleksandr Pilgun}, {and} \bibinfo{person}{Yury
  Zhauniarovich}.} \bibinfo{year}{2018}\natexlab{}.
\newblock \showarticletitle{The Influence of Code Coverage Metrics on Automated
  Testing Efficiency in Android}. In \bibinfo{booktitle}{\emph{Proceedings of
  the 2018 ACM SIGSAC Conference on Computer and Communications Security}}
  (Toronto, Canada) \emph{(\bibinfo{series}{CCS '18})}.
  \bibinfo{publisher}{Association for Computing Machinery},
  \bibinfo{address}{New York, NY, USA}, \bibinfo{pages}{2216–2218}.
\newblock
\showISBNx{9781450356930}
\urldef\tempurl%
\url{https://doi.org/10.1145/3243734.3278524}
\showDOI{\tempurl}


\bibitem[Dong et~al\mbox{.}(2020)]%
        {dong2020time}
\bibfield{author}{\bibinfo{person}{Zhen Dong}, \bibinfo{person}{Marcel
  B{\"o}hme}, \bibinfo{person}{Lucia Cojocaru}, {and} \bibinfo{person}{Abhik
  Roychoudhury}.} \bibinfo{year}{2020}\natexlab{}.
\newblock \showarticletitle{Time-travel testing of android apps}. In
  \bibinfo{booktitle}{\emph{2020 IEEE/ACM 42nd International Conference on
  Software Engineering (ICSE)}}. IEEE, \bibinfo{pages}{481--492}.
\newblock


\bibitem[El~Zarif et~al\mbox{.}(2020)]%
        {el2020relationship}
\bibfield{author}{\bibinfo{person}{Omar El~Zarif},
  \bibinfo{person}{Daniel~Alencar Da~Costa}, \bibinfo{person}{Safwat Hassan},
  {and} \bibinfo{person}{Ying Zou}.} \bibinfo{year}{2020}\natexlab{}.
\newblock \showarticletitle{On the relationship between user churn and software
  issues}. In \bibinfo{booktitle}{\emph{Proceedings of the 17th International
  Conference on Mining Software Repositories}}. \bibinfo{pages}{339--349}.
\newblock


\bibitem[Feng et~al\mbox{.}(2023)]%
        {feng2023efficiency}
\bibfield{author}{\bibinfo{person}{Sidong Feng}, \bibinfo{person}{Mulong Xie},
  {and} \bibinfo{person}{Chunyang Chen}.} \bibinfo{year}{2023}\natexlab{}.
\newblock \showarticletitle{Efficiency matters: Speeding up automated testing
  with gui rendering inference}. In \bibinfo{booktitle}{\emph{2023 IEEE/ACM
  45th International Conference on Software Engineering (ICSE)}}. IEEE,
  \bibinfo{pages}{906--918}.
\newblock


\bibitem[Gao et~al\mbox{.}(2018)]%
        {gao2018android}
\bibfield{author}{\bibinfo{person}{Xiang Gao}, \bibinfo{person}{Shin~Hwei Tan},
  \bibinfo{person}{Zhen Dong}, {and} \bibinfo{person}{Abhik Roychoudhury}.}
  \bibinfo{year}{2018}\natexlab{}.
\newblock \showarticletitle{Android testing via synthetic symbolic execution}.
  In \bibinfo{booktitle}{\emph{2018 33rd IEEE/ACM International Conference on
  Automated Software Engineering (ASE)}}. IEEE, \bibinfo{pages}{419--429}.
\newblock


\bibitem[Gu et~al\mbox{.}(2017)]%
        {gu2017aimdroid}
\bibfield{author}{\bibinfo{person}{Tianxiao Gu}, \bibinfo{person}{Chun Cao},
  \bibinfo{person}{Tianchi Liu}, \bibinfo{person}{Chengnian Sun},
  \bibinfo{person}{Jing Deng}, \bibinfo{person}{Xiaoxing Ma}, {and}
  \bibinfo{person}{Jian L{\"u}}.} \bibinfo{year}{2017}\natexlab{}.
\newblock \showarticletitle{Aimdroid: Activity-insulated multi-level automated
  testing for android applications}. In \bibinfo{booktitle}{\emph{2017 IEEE
  International Conference on Software Maintenance and Evolution (ICSME)}}.
  IEEE, \bibinfo{pages}{103--114}.
\newblock


\bibitem[Gu et~al\mbox{.}(2019)]%
        {gu2019practical}
\bibfield{author}{\bibinfo{person}{Tianxiao Gu}, \bibinfo{person}{Chengnian
  Sun}, \bibinfo{person}{Xiaoxing Ma}, \bibinfo{person}{Chun Cao},
  \bibinfo{person}{Chang Xu}, \bibinfo{person}{Yuan Yao},
  \bibinfo{person}{Qirun Zhang}, \bibinfo{person}{Jian Lu}, {and}
  \bibinfo{person}{Zhendong Su}.} \bibinfo{year}{2019}\natexlab{}.
\newblock \showarticletitle{Practical GUI testing of Android applications via
  model abstraction and refinement}. In \bibinfo{booktitle}{\emph{2019 IEEE/ACM
  41st International Conference on Software Engineering (ICSE)}}. IEEE,
  \bibinfo{pages}{269--280}.
\newblock


\bibitem[Guo et~al\mbox{.}(2020)]%
        {guo2020improving}
\bibfield{author}{\bibinfo{person}{Wunan Guo}, \bibinfo{person}{Liwei Shen},
  \bibinfo{person}{Ting Su}, \bibinfo{person}{Xin Peng}, {and}
  \bibinfo{person}{Weiyang Xie}.} \bibinfo{year}{2020}\natexlab{}.
\newblock \showarticletitle{Improving automated GUI exploration of android apps
  via static dependency analysis}. In \bibinfo{booktitle}{\emph{2020 IEEE
  International Conference on Software Maintenance and Evolution (ICSME)}}.
  IEEE, \bibinfo{pages}{557--568}.
\newblock


\bibitem[Hoffmann et~al\mbox{.}(2013)]%
        {hoffmann2013slicing}
\bibfield{author}{\bibinfo{person}{Johannes Hoffmann}, \bibinfo{person}{Martin
  Ussath}, \bibinfo{person}{Thorsten Holz}, {and} \bibinfo{person}{Michael
  Spreitzenbarth}.} \bibinfo{year}{2013}\natexlab{}.
\newblock \showarticletitle{Slicing droids: program slicing for smali code}. In
  \bibinfo{booktitle}{\emph{Proceedings of the 28th Annual ACM Symposium on
  Applied Computing}}. \bibinfo{pages}{1844--1851}.
\newblock


\bibitem[Hu and Neamtiu(2011)]%
        {hu2011automating}
\bibfield{author}{\bibinfo{person}{Cuixiong Hu} {and} \bibinfo{person}{Iulian
  Neamtiu}.} \bibinfo{year}{2011}\natexlab{}.
\newblock \showarticletitle{Automating GUI testing for Android applications}.
  In \bibinfo{booktitle}{\emph{Proceedings of the 6th International Workshop on
  Automation of Software Test}}. \bibinfo{pages}{77--83}.
\newblock


\bibitem[Hu et~al\mbox{.}(2014)]%
        {hu2014efficiently}
\bibfield{author}{\bibinfo{person}{Gang Hu}, \bibinfo{person}{Xinhao Yuan},
  \bibinfo{person}{Yang Tang}, {and} \bibinfo{person}{Junfeng Yang}.}
  \bibinfo{year}{2014}\natexlab{}.
\newblock \showarticletitle{Efficiently, effectively detecting mobile app bugs
  with appdoctor}. In \bibinfo{booktitle}{\emph{Proceedings of the Ninth
  European Conference on Computer Systems}}. \bibinfo{pages}{1--15}.
\newblock


\bibitem[Hu et~al\mbox{.}(2019)]%
        {hu2019code}
\bibfield{author}{\bibinfo{person}{Han Hu}, \bibinfo{person}{Qiuyuan Chen},
  {and} \bibinfo{person}{Zhaoyi Liu}.} \bibinfo{year}{2019}\natexlab{}.
\newblock \showarticletitle{Code generation from supervised code embeddings}.
  In \bibinfo{booktitle}{\emph{Neural Information Processing: 26th
  International Conference, ICONIP 2019, Sydney, NSW, Australia, December
  12--15, 2019, Proceedings, Part IV 26}}. Springer, \bibinfo{pages}{388--396}.
\newblock


\bibitem[Hu et~al\mbox{.}(2023a)]%
        {hu2024TVGUI}
\bibfield{author}{\bibinfo{person}{Han Hu}, \bibinfo{person}{Ruiqi Dong},
  \bibinfo{person}{John Grundy}, \bibinfo{person}{Thai~Minh Nguyen},
  \bibinfo{person}{Huaxiao Liu}, {and} \bibinfo{person}{Chunyang Chen}.}
  \bibinfo{year}{2023}\natexlab{a}.
\newblock \showarticletitle{Automated Mapping of Adaptive App GUIs from Phones
  to TVs}.
\newblock \bibinfo{journal}{\emph{ACM Trans. Softw. Eng. Methodol.}}
  \bibinfo{volume}{33}, \bibinfo{number}{2}, Article \bibinfo{articleno}{47}
  (\bibinfo{date}{dec} \bibinfo{year}{2023}), \bibinfo{numpages}{31}~pages.
\newblock
\showISSN{1049-331X}
\urldef\tempurl%
\url{https://doi.org/10.1145/3631968}
\showDOI{\tempurl}


\bibitem[Hu et~al\mbox{.}(2023b)]%
        {hu2024iOS}
\bibfield{author}{\bibinfo{person}{Han Hu}, \bibinfo{person}{Yujin Huang},
  \bibinfo{person}{Qiuyuan Chen}, \bibinfo{person}{Terry~Yue Zhuo}, {and}
  \bibinfo{person}{Chunyang Chen}.} \bibinfo{year}{2023}\natexlab{b}.
\newblock \showarticletitle{A First Look at On-device Models in iOS Apps}.
\newblock \bibinfo{journal}{\emph{ACM Trans. Softw. Eng. Methodol.}}
  \bibinfo{volume}{33}, \bibinfo{number}{1}, Article \bibinfo{articleno}{26}
  (\bibinfo{date}{nov} \bibinfo{year}{2023}), \bibinfo{numpages}{30}~pages.
\newblock
\showISSN{1049-331X}
\urldef\tempurl%
\url{https://doi.org/10.1145/3617177}
\showDOI{\tempurl}


\bibitem[Hu et~al\mbox{.}(2023c)]%
        {hu2023pairwise}
\bibfield{author}{\bibinfo{person}{Han Hu}, \bibinfo{person}{Haolan Zhan},
  \bibinfo{person}{Yujin Huang}, {and} \bibinfo{person}{Di Liu}.}
  \bibinfo{year}{2023}\natexlab{c}.
\newblock \showarticletitle{Pairwise GUI Dataset Construction Between Android
  Phones and Tablets}.
\newblock \bibinfo{journal}{\emph{arXiv preprint arXiv:2310.04755}}
  (\bibinfo{year}{2023}).
\newblock


\bibitem[Hu and Liu(2024)]%
        {hu2024pairwise}
\bibfield{author}{\bibinfo{person}{Zhan Haolan Huang~Yujin Hu, Han} {and}
  \bibinfo{person}{Di Liu}.} \bibinfo{year}{2024}\natexlab{}.
\newblock \showarticletitle{Pairwise GUI dataset construction between Android
  phones and tablets}.
\newblock \bibinfo{journal}{\emph{Advances in Neural Information Processing
  Systems}}  \bibinfo{volume}{36} (\bibinfo{year}{2024}).
\newblock


\bibitem[Jensen et~al\mbox{.}(2013)]%
        {jensen2013automated}
\bibfield{author}{\bibinfo{person}{Casper~S Jensen}, \bibinfo{person}{Mukul~R
  Prasad}, {and} \bibinfo{person}{Anders M{\o}ller}.}
  \bibinfo{year}{2013}\natexlab{}.
\newblock \showarticletitle{Automated testing with targeted event sequence
  generation}. In \bibinfo{booktitle}{\emph{Proceedings of the 2013
  International Symposium on Software Testing and Analysis}}.
  \bibinfo{pages}{67--77}.
\newblock


\bibitem[Kowalczyk et~al\mbox{.}(2018)]%
        {kowalczyk2018configurations}
\bibfield{author}{\bibinfo{person}{Emily Kowalczyk}, \bibinfo{person}{Myra~B
  Cohen}, {and} \bibinfo{person}{Atif~M Memon}.}
  \bibinfo{year}{2018}\natexlab{}.
\newblock \showarticletitle{Configurations in Android testing: they matter}. In
  \bibinfo{booktitle}{\emph{Proceedings of the 1st International Workshop on
  Advances in Mobile App Analysis}}. \bibinfo{pages}{1--6}.
\newblock


\bibitem[Lai and Rubin(2019)]%
        {lai2019goal}
\bibfield{author}{\bibinfo{person}{Duling Lai} {and} \bibinfo{person}{Julia
  Rubin}.} \bibinfo{year}{2019}\natexlab{}.
\newblock \showarticletitle{Goal-driven exploration for android applications}.
  In \bibinfo{booktitle}{\emph{2019 34th IEEE/ACM International Conference on
  Automated Software Engineering (ASE)}}. IEEE, \bibinfo{pages}{115--127}.
\newblock


\bibitem[Li et~al\mbox{.}(2019)]%
        {li2019humanoid}
\bibfield{author}{\bibinfo{person}{Yuanchun Li}, \bibinfo{person}{Ziyue Yang},
  \bibinfo{person}{Yao Guo}, {and} \bibinfo{person}{Xiangqun Chen}.}
  \bibinfo{year}{2019}\natexlab{}.
\newblock \showarticletitle{Humanoid: a deep learning-based approach to
  automated black-box Android app testing}. In \bibinfo{booktitle}{\emph{2019
  34th IEEE/ACM International Conference on Automated Software Engineering
  (ASE)}}. IEEE, \bibinfo{pages}{1070--1073}.
\newblock


\bibitem[Liu et~al\mbox{.}(2022)]%
        {liu2022dalt}
\bibfield{author}{\bibinfo{person}{Ao Liu}, \bibinfo{person}{Chenkai Guo},
  \bibinfo{person}{Naipeng Dong}, \bibinfo{person}{Yinjie Wang}, {and}
  \bibinfo{person}{Jing Xu}.} \bibinfo{year}{2022}\natexlab{}.
\newblock \showarticletitle{DALT: Deep Activity Launching Test via
  Intent-Constraint Extraction}. In \bibinfo{booktitle}{\emph{2022 IEEE 33rd
  International Symposium on Software Reliability Engineering (ISSRE)}}. IEEE,
  \bibinfo{pages}{482--493}.
\newblock


\bibitem[Liu and Xiao(2023)]%
        {liu2021promal}
\bibfield{author}{\bibinfo{person}{Changlin Liu} {and} \bibinfo{person}{Xusheng
  Xiao}.} \bibinfo{year}{2023}\natexlab{}.
\newblock \showarticletitle{ProMal: precise window transition graphs for
  Android via synergy of program analysis and machine learning}. In
  \bibinfo{booktitle}{\emph{International Conference on Software Engineering}}.
  IEEE.
\newblock


\bibitem[Liu et~al\mbox{.}(2023)]%
        {liu2023fill}
\bibfield{author}{\bibinfo{person}{Zhe Liu}, \bibinfo{person}{Chunyang Chen},
  \bibinfo{person}{Junjie Wang}, \bibinfo{person}{Xing Che},
  \bibinfo{person}{Yuekai Huang}, \bibinfo{person}{Jun Hu}, {and}
  \bibinfo{person}{Qing Wang}.} \bibinfo{year}{2023}\natexlab{}.
\newblock \showarticletitle{Fill in the blank: Context-aware automated text
  input generation for mobile gui testing}. In \bibinfo{booktitle}{\emph{2023
  IEEE/ACM 45th International Conference on Software Engineering (ICSE)}}.
  IEEE, \bibinfo{pages}{1355--1367}.
\newblock


\bibitem[Liu et~al\mbox{.}(2024)]%
        {liu2024testing}
\bibfield{author}{\bibinfo{person}{Zhe Liu}, \bibinfo{person}{Chunyang Chen},
  \bibinfo{person}{Junjie Wang}, \bibinfo{person}{Mengzhuo Chen},
  \bibinfo{person}{Boyu Wu}, \bibinfo{person}{Zhilin Tian},
  \bibinfo{person}{Yuekai Huang}, \bibinfo{person}{Jun Hu}, {and}
  \bibinfo{person}{Qing Wang}.} \bibinfo{year}{2024}\natexlab{}.
\newblock \showarticletitle{Testing the limits: Unusual text inputs generation
  for mobile app crash detection with large language model}. In
  \bibinfo{booktitle}{\emph{Proceedings of the IEEE/ACM 46th International
  Conference on Software Engineering}}. \bibinfo{pages}{1--12}.
\newblock


\bibitem[Lv et~al\mbox{.}(2022)]%
        {lv2022fastbot2}
\bibfield{author}{\bibinfo{person}{Zhengwei Lv}, \bibinfo{person}{Chao Peng},
  \bibinfo{person}{Zhao Zhang}, \bibinfo{person}{Ting Su}, \bibinfo{person}{Kai
  Liu}, {and} \bibinfo{person}{Ping Yang}.} \bibinfo{year}{2022}\natexlab{}.
\newblock \showarticletitle{Fastbot2: Reusable Automated Model-based GUI
  Testing for Android Enhanced by Reinforcement Learning}. In
  \bibinfo{booktitle}{\emph{37th IEEE/ACM International Conference on Automated
  Software Engineering}}. \bibinfo{pages}{1--5}.
\newblock


\bibitem[Ma et~al\mbox{.}(2018)]%
        {ma2018aladdin}
\bibfield{author}{\bibinfo{person}{Yun Ma}, \bibinfo{person}{Ziniu Hu},
  \bibinfo{person}{Yunxin Liu}, \bibinfo{person}{Tao Xie}, {and}
  \bibinfo{person}{Xuanzhe Liu}.} \bibinfo{year}{2018}\natexlab{}.
\newblock \showarticletitle{Aladdin: Automating release of deep-link APIs on
  Android}. In \bibinfo{booktitle}{\emph{Proceedings of the 2018 World Wide Web
  Conference}}. \bibinfo{pages}{1469--1478}.
\newblock


\bibitem[Machiry et~al\mbox{.}(2013)]%
        {machiry2013dynodroid}
\bibfield{author}{\bibinfo{person}{Aravind Machiry}, \bibinfo{person}{Rohan
  Tahiliani}, {and} \bibinfo{person}{Mayur Naik}.}
  \bibinfo{year}{2013}\natexlab{}.
\newblock \showarticletitle{Dynodroid: An input generation system for android
  apps}. In \bibinfo{booktitle}{\emph{Proceedings of the 2013 9th Joint Meeting
  on Foundations of Software Engineering}}. \bibinfo{pages}{224--234}.
\newblock


\bibitem[Mahmood et~al\mbox{.}(2012)]%
        {mahmood2012whitebox}
\bibfield{author}{\bibinfo{person}{Riyadh Mahmood}, \bibinfo{person}{Naeem
  Esfahani}, \bibinfo{person}{Thabet Kacem}, \bibinfo{person}{Nariman Mirzaei},
  \bibinfo{person}{Sam Malek}, {and} \bibinfo{person}{Angelos Stavrou}.}
  \bibinfo{year}{2012}\natexlab{}.
\newblock \showarticletitle{A whitebox approach for automated security testing
  of Android applications on the cloud}. In \bibinfo{booktitle}{\emph{2012 7th
  International Workshop on Automation of Software Test (AST)}}. IEEE,
  \bibinfo{pages}{22--28}.
\newblock


\bibitem[Mao et~al\mbox{.}(2016)]%
        {mao2016sapienz}
\bibfield{author}{\bibinfo{person}{Ke Mao}, \bibinfo{person}{Mark Harman},
  {and} \bibinfo{person}{Yue Jia}.} \bibinfo{year}{2016}\natexlab{}.
\newblock \showarticletitle{Sapienz: Multi-objective automated testing for
  android applications}. In \bibinfo{booktitle}{\emph{Proceedings of the 25th
  international symposium on software testing and analysis}}.
  \bibinfo{pages}{94--105}.
\newblock


\bibitem[Mao et~al\mbox{.}(2017)]%
        {mao2017crowd}
\bibfield{author}{\bibinfo{person}{Ke Mao}, \bibinfo{person}{Mark Harman},
  {and} \bibinfo{person}{Yue Jia}.} \bibinfo{year}{2017}\natexlab{}.
\newblock \showarticletitle{Crowd intelligence enhances automated mobile
  testing}. In \bibinfo{booktitle}{\emph{2017 32nd IEEE/ACM International
  Conference on Automated Software Engineering (ASE)}}. IEEE,
  \bibinfo{pages}{16--26}.
\newblock


\bibitem[McKnight and Najab(2010)]%
        {mcknight2010mann}
\bibfield{author}{\bibinfo{person}{Patrick~E McKnight} {and}
  \bibinfo{person}{Julius Najab}.} \bibinfo{year}{2010}\natexlab{}.
\newblock \showarticletitle{Mann-Whitney U Test}.
\newblock \bibinfo{journal}{\emph{The Corsini encyclopedia of psychology}}
  (\bibinfo{year}{2010}), \bibinfo{pages}{1--1}.
\newblock


\bibitem[Milanova et~al\mbox{.}(2005)]%
        {milanova2005parameterized}
\bibfield{author}{\bibinfo{person}{Ana Milanova}, \bibinfo{person}{Atanas
  Rountev}, {and} \bibinfo{person}{Barbara~G Ryder}.}
  \bibinfo{year}{2005}\natexlab{}.
\newblock \showarticletitle{Parameterized object sensitivity for points-to
  analysis for Java}.
\newblock \bibinfo{journal}{\emph{ACM Transactions on Software Engineering and
  Methodology (TOSEM)}} \bibinfo{volume}{14}, \bibinfo{number}{1}
  (\bibinfo{year}{2005}), \bibinfo{pages}{1--41}.
\newblock


\bibitem[Mirzaei et~al\mbox{.}(2012)]%
        {mirzaei2012testing}
\bibfield{author}{\bibinfo{person}{Nariman Mirzaei}, \bibinfo{person}{Sam
  Malek}, \bibinfo{person}{Corina~S P{\u{a}}s{\u{a}}reanu},
  \bibinfo{person}{Naeem Esfahani}, {and} \bibinfo{person}{Riyadh Mahmood}.}
  \bibinfo{year}{2012}\natexlab{}.
\newblock \showarticletitle{Testing android apps through symbolic execution}.
\newblock \bibinfo{journal}{\emph{ACM SIGSOFT Software Engineering Notes}}
  \bibinfo{volume}{37}, \bibinfo{number}{6} (\bibinfo{year}{2012}),
  \bibinfo{pages}{1--5}.
\newblock


\bibitem[Octeau et~al\mbox{.}(2015)]%
        {octeau2015composite}
\bibfield{author}{\bibinfo{person}{Damien Octeau}, \bibinfo{person}{Daniel
  Luchaup}, \bibinfo{person}{Matthew Dering}, \bibinfo{person}{Somesh Jha},
  {and} \bibinfo{person}{Patrick McDaniel}.} \bibinfo{year}{2015}\natexlab{}.
\newblock \showarticletitle{Composite constant propagation: Application to
  android inter-component communication analysis}. In
  \bibinfo{booktitle}{\emph{2015 IEEE/ACM 37th IEEE International Conference on
  Software Engineering}}, Vol.~\bibinfo{volume}{1}. IEEE,
  \bibinfo{pages}{77--88}.
\newblock


\bibitem[Octeau et~al\mbox{.}(2013)]%
        {octeau2013effective}
\bibfield{author}{\bibinfo{person}{Damien Octeau}, \bibinfo{person}{Patrick
  McDaniel}, \bibinfo{person}{Somesh Jha}, \bibinfo{person}{Alexandre Bartel},
  \bibinfo{person}{Eric Bodden}, \bibinfo{person}{Jacques Klein}, {and}
  \bibinfo{person}{Yves Le~Traon}.} \bibinfo{year}{2013}\natexlab{}.
\newblock \showarticletitle{Effective $\{$Inter-Component$\}$ Communication
  Mapping in Android: An Essential Step Towards Holistic Security Analysis}. In
  \bibinfo{booktitle}{\emph{22nd USENIX Security Symposium (USENIX Security
  13)}}. \bibinfo{pages}{543--558}.
\newblock


\bibitem[Pan et~al\mbox{.}(2020)]%
        {pan2020reinforcement}
\bibfield{author}{\bibinfo{person}{Minxue Pan}, \bibinfo{person}{An Huang},
  \bibinfo{person}{Guoxin Wang}, \bibinfo{person}{Tian Zhang}, {and}
  \bibinfo{person}{Xuandong Li}.} \bibinfo{year}{2020}\natexlab{}.
\newblock \showarticletitle{Reinforcement learning based curiosity-driven
  testing of Android applications}. In \bibinfo{booktitle}{\emph{Proceedings of
  the 29th ACM SIGSOFT International Symposium on Software Testing and
  Analysis}}. \bibinfo{pages}{153--164}.
\newblock


\bibitem[Planning(2002)]%
        {planning2002economic}
\bibfield{author}{\bibinfo{person}{Strategic Planning}.}
  \bibinfo{year}{2002}\natexlab{}.
\newblock \showarticletitle{The economic impacts of inadequate infrastructure
  for software testing}.
\newblock \bibinfo{journal}{\emph{National Institute of Standards and
  Technology}}  \bibinfo{volume}{1} (\bibinfo{year}{2002}).
\newblock


\bibitem[Su et~al\mbox{.}(2017)]%
        {su2017guided}
\bibfield{author}{\bibinfo{person}{Ting Su}, \bibinfo{person}{Guozhu Meng},
  \bibinfo{person}{Yuting Chen}, \bibinfo{person}{Ke Wu},
  \bibinfo{person}{Weiming Yang}, \bibinfo{person}{Yao Yao},
  \bibinfo{person}{Geguang Pu}, \bibinfo{person}{Yang Liu}, {and}
  \bibinfo{person}{Zhendong Su}.} \bibinfo{year}{2017}\natexlab{}.
\newblock \showarticletitle{Guided, stochastic model-based GUI testing of
  Android apps}. In \bibinfo{booktitle}{\emph{Proceedings of the 2017 11th
  Joint Meeting on Foundations of Software Engineering}}.
  \bibinfo{pages}{245--256}.
\newblock


\bibitem[Su et~al\mbox{.}(2021a)]%
        {su2021benchmarking}
\bibfield{author}{\bibinfo{person}{Ting Su}, \bibinfo{person}{Jue Wang}, {and}
  \bibinfo{person}{Zhendong Su}.} \bibinfo{year}{2021}\natexlab{a}.
\newblock \showarticletitle{Benchmarking automated gui testing for android
  against real-world bugs}. In \bibinfo{booktitle}{\emph{Proceedings of the
  29th ACM Joint Meeting on European Software Engineering Conference and
  Symposium on the Foundations of Software Engineering}}.
  \bibinfo{pages}{119--130}.
\newblock


\bibitem[Su et~al\mbox{.}(2021b)]%
        {su2021fully}
\bibfield{author}{\bibinfo{person}{Ting Su}, \bibinfo{person}{Yichen Yan},
  \bibinfo{person}{Jue Wang}, \bibinfo{person}{Jingling Sun},
  \bibinfo{person}{Yiheng Xiong}, \bibinfo{person}{Geguang Pu},
  \bibinfo{person}{Ke Wang}, {and} \bibinfo{person}{Zhendong Su}.}
  \bibinfo{year}{2021}\natexlab{b}.
\newblock \showarticletitle{Fully automated functional fuzzing of Android apps
  for detecting non-crashing logic bugs}.
\newblock \bibinfo{journal}{\emph{Proceedings of the ACM on Programming
  Languages}} \bibinfo{volume}{5}, \bibinfo{number}{OOPSLA}
  (\bibinfo{year}{2021}), \bibinfo{pages}{1--31}.
\newblock


\bibitem[Tang et~al\mbox{.}(2020)]%
        {tang2020all}
\bibfield{author}{\bibinfo{person}{Yutian Tang}, \bibinfo{person}{Yulei Sui},
  \bibinfo{person}{Haoyu Wang}, \bibinfo{person}{Xiapu Luo},
  \bibinfo{person}{Hao Zhou}, {and} \bibinfo{person}{Zhou Xu}.}
  \bibinfo{year}{2020}\natexlab{}.
\newblock \showarticletitle{All your app links are belong to us: understanding
  the threats of instant apps based attacks}. In
  \bibinfo{booktitle}{\emph{Proceedings of the 28th ACM Joint Meeting on
  European Software Engineering Conference and Symposium on the Foundations of
  Software Engineering}}. \bibinfo{pages}{914--926}.
\newblock


\bibitem[Tian et~al\mbox{.}(2018)]%
        {tian2018poster}
\bibfield{author}{\bibinfo{person}{Cong Tian}, \bibinfo{person}{Congli Xia},
  {and} \bibinfo{person}{Zhenhua Duan}.} \bibinfo{year}{2018}\natexlab{}.
\newblock \showarticletitle{Poster: Android Inter-Component Communication
  Analysis with Intent Revision}. In \bibinfo{booktitle}{\emph{2018 IEEE/ACM
  40th International Conference on Software Engineering: Companion
  (ICSE-Companion)}}. IEEE, \bibinfo{pages}{254--255}.
\newblock


\bibitem[Tian et~al\mbox{.}(2015)]%
        {tian2015characteristics}
\bibfield{author}{\bibinfo{person}{Yuan Tian}, \bibinfo{person}{Meiyappan
  Nagappan}, \bibinfo{person}{David Lo}, {and} \bibinfo{person}{Ahmed~E
  Hassan}.} \bibinfo{year}{2015}\natexlab{}.
\newblock \showarticletitle{What are the characteristics of high-rated apps? a
  case study on free android applications}. In \bibinfo{booktitle}{\emph{2015
  IEEE international conference on software maintenance and evolution
  (ICSME)}}. IEEE, \bibinfo{pages}{301--310}.
\newblock


\bibitem[Wang et~al\mbox{.}(2022)]%
        {wang2022detecting}
\bibfield{author}{\bibinfo{person}{Jue Wang}, \bibinfo{person}{Yanyan Jiang},
  \bibinfo{person}{Ting Su}, \bibinfo{person}{Shaohua Li},
  \bibinfo{person}{Chang Xu}, \bibinfo{person}{Jian Lu}, {and}
  \bibinfo{person}{Zhendong Su}.} \bibinfo{year}{2022}\natexlab{}.
\newblock \showarticletitle{Detecting non-crashing functional bugs in Android
  apps via deep-state differential analysis}. In
  \bibinfo{booktitle}{\emph{Proceedings of the 30th ACM Joint European Software
  Engineering Conference and Symposium on the Foundations of Software
  Engineering}}. \bibinfo{pages}{434--446}.
\newblock


\bibitem[Wang et~al\mbox{.}(2018)]%
        {wang2018empirical}
\bibfield{author}{\bibinfo{person}{Wenyu Wang}, \bibinfo{person}{Dengfeng Li},
  \bibinfo{person}{Wei Yang}, \bibinfo{person}{Yurui Cao},
  \bibinfo{person}{Zhenwen Zhang}, \bibinfo{person}{Yuetang Deng}, {and}
  \bibinfo{person}{Tao Xie}.} \bibinfo{year}{2018}\natexlab{}.
\newblock \showarticletitle{An empirical study of android test generation tools
  in industrial cases}. In \bibinfo{booktitle}{\emph{2018 33rd IEEE/ACM
  International Conference on Automated Software Engineering (ASE)}}. IEEE,
  \bibinfo{pages}{738--748}.
\newblock


\bibitem[Wang et~al\mbox{.}(2021)]%
        {wang2021vet}
\bibfield{author}{\bibinfo{person}{Wenyu Wang}, \bibinfo{person}{Wei Yang},
  \bibinfo{person}{Tianyin Xu}, {and} \bibinfo{person}{Tao Xie}.}
  \bibinfo{year}{2021}\natexlab{}.
\newblock \showarticletitle{Vet: identifying and avoiding UI exploration
  tarpits}. In \bibinfo{booktitle}{\emph{Proceedings of the 29th ACM Joint
  Meeting on European Software Engineering Conference and Symposium on the
  Foundations of Software Engineering}}. \bibinfo{pages}{83--94}.
\newblock


\bibitem[Yan et~al\mbox{.}(2020)]%
        {yan2020multiple}
\bibfield{author}{\bibinfo{person}{Jiwei Yan}, \bibinfo{person}{Hao Liu},
  \bibinfo{person}{Linjie Pan}, \bibinfo{person}{Jun Yan},
  \bibinfo{person}{Jian Zhang}, {and} \bibinfo{person}{Bin Liang}.}
  \bibinfo{year}{2020}\natexlab{}.
\newblock \showarticletitle{Multiple-entry testing of android applications by
  constructing activity launching contexts}. In \bibinfo{booktitle}{\emph{2020
  IEEE/ACM 42nd International Conference on Software Engineering (ICSE)}}.
  IEEE, \bibinfo{pages}{457--468}.
\newblock


\bibitem[Yang et~al\mbox{.}(2018)]%
        {yang2018static}
\bibfield{author}{\bibinfo{person}{Shengqian Yang}, \bibinfo{person}{Haowei
  Wu}, \bibinfo{person}{Hailong Zhang}, \bibinfo{person}{Yan Wang},
  \bibinfo{person}{Chandrasekar Swaminathan}, \bibinfo{person}{Dacong Yan},
  {and} \bibinfo{person}{Atanas Rountev}.} \bibinfo{year}{2018}\natexlab{}.
\newblock \showarticletitle{Static window transition graphs for Android}.
\newblock \bibinfo{journal}{\emph{Automated Software Engineering}}
  \bibinfo{volume}{25}, \bibinfo{number}{4} (\bibinfo{year}{2018}),
  \bibinfo{pages}{833--873}.
\newblock


\bibitem[Zhengwei et~al\mbox{.}(2022)]%
        {fastbot2}
\bibfield{author}{\bibinfo{person}{Lv Zhengwei}, \bibinfo{person}{Peng Chao},
  \bibinfo{person}{Zhang Zhao}, \bibinfo{person}{Su Ting}, \bibinfo{person}{Liu
  Kai}, {and} \bibinfo{person}{Yang Ping}.} \bibinfo{year}{2022}\natexlab{}.
\newblock \showarticletitle{Fastbot2: Reusable Automated Model-based GUI
  Testing for Android Enhanced by Reinforcement Learning}. In
  \bibinfo{booktitle}{\emph{37th IEEE/ACM International Conference on Automated
  Software Engineering (ASE)}}.
\newblock


\end{thebibliography}

\end{document}